\documentclass[sigconf]{acmart}

\usepackage{subfig}
\usepackage{url}

\usepackage{multicol}
\usepackage{multirow}
\usepackage{colortbl}
\usepackage{xspace}
\usepackage{pgfplots}
\pgfplotsset{compat=1.18}
\definecolor{mygray}{HTML}{ededed}
\usepackage{longtable}

\usepackage{caption}
\usepackage{float}

\usepackage{tikz}
\usetikzlibrary{shapes.geometric, arrows, positioning}

\tikzstyle{process} = [rectangle, minimum width=4cm, minimum height=1cm, text centered, text width=4cm, draw=black, fill=white, rounded corners]
\tikzstyle{arrow} = [thick,->,>=stealth]
\tikzstyle{label} = [text centered, fill=blue!30, text width=2cm, font=\bfseries]

\usepackage{booktabs}

\newcommand{\verttext}[1]{
  \parbox[t]{2mm}{\rotatebox[origin=lB]{60}{#1}}
}
\newcommand{\yes}{$\medbullet$\xspace}
\newcommand{\no}{$-$}
\newcommand{\leftrule}[1]{
    \multicolumn{1}{|c}{#1}
}

\setcopyright{acmlicensed}
\copyrightyear{2026}

\acmYear{2026}\copyrightyear{2026}
\setcopyright{cc}
\setcctype[4.0]{by}
\acmConference[ASIA CCS '26]{ACM Asia Conference on Computer and Communications Security}{June 1--5, 2026}{Bangalore, India}
\acmBooktitle{ACM Asia Conference on Computer and Communications Security (ASIA CCS '26), June 1--5, 2026, Bangalore, India}
\acmDOI{10.1145/3779208.3785270}
\acmISBN{979-8-4007-2356-8/26/06}

\author{Suleiman Saka}
\affiliation{%
 \institution{University of Denver}
 \city{Denver}
 \state{Colorado}
 \country{USA}}
 \email{suleiman.saka@du.edu}

\author{Sanchari Das}
\affiliation{%
  \institution{George Mason University}
  \city{Fairfax}
  \state{Virginia}
  \country{USA}}
\email{sdas35@gmu.edu}

\renewcommand{\shortauthors}{Saka and Das}

\begin{document}

\title[SoK: IoT Security, Privacy, Accessibility, and Usability for Older Adults]{SoK: Reviewing Two Decades of Security, Privacy, Accessibility, and Usability Studies on Internet of Things for Older Adults}

\begin{abstract}
The Internet of Things (IoT) has the potential to enhance older adults’ independence and quality of life, but it also exposes them to security, privacy, accessibility, and usability (SPAU) risks. We conducted a systematic review of $44$ peer-reviewed studies published between $2004$ and $2024$ using a five-phase screening pipeline. From each study, we extracted data on study design, IoT type, SPAU measures, and identified research gaps. We introduce the~\emph{SPAU-IoT Framework}, which comprises $27$ criteria across four dimensions: security (e.g., resilience to cyber threats, secure authentication, encrypted communication, secure-by-default settings, and guardianship features), privacy (e.g., data minimization, explicit consent, and privacy-preserving analytics), accessibility (e.g., compliance with ADA/WCAG standards and assistive-technology compatibility), and usability (e.g., guided interaction, integrated assistance, and progressive learning). Applying this framework revealed that more than $70\%$ of studies implemented authentication and encryption mechanisms, whereas fewer than $50\%$ addressed accessibility or usability concerns. We further developed a threat model that maps IoT assets, networks, and backend servers to exploit vectors such as phishing, caregiver exploitation, and weak-password attacks, explicitly accounting for age-related vulnerabilities including cognitive decline and sensory impairment. Our results expose a systemic lack of integrated SPAU approaches in existing IoT research and translate these gaps into actionable, standards-aligned design guidelines for IoT systems designed for older adults.
\end{abstract}





\maketitle

\section{Introduction}
The World Health Organization (WHO) projects that by $2030$, the global population aged $65+$ will reach $1.4$ billion, rising to $2.1$ billion by $2050$~\cite{WHO_ageing}. This demographic shift poses significant challenges in meeting the needs of an aging population~\cite{dobre_improving_2019, KUMAR2023110720}. The Internet of Things (IoT) offers promising avenues, such as remote monitoring, assistive technologies, and smart homes to enhance older adults' independence and quality of life~\cite{das2022leveraging,saka2023safeguarding}. Yet, IoT adoption reveals critical gaps in addressing security, privacy, accessibility, and usability (SPAU) concerns~\cite{de_carli_vision_2021,toutsop2021exploring,tazi2023accessibility,hadan2019making,neupane2022data,tazi2025multi,gopavaram2019iotmarketplace}. Age-related physical, cognitive, sensory, and technological limitations are often overlooked in IoT design~\cite{saka2025watch}. For example, reduced motor abilities and arthritis can hinder interaction with small devices or touchscreens~\cite{Wu_BridgingTD, Tanja_2023_OA}; cognitive decline impairs navigation of complex interfaces, password recall, and security management~\cite{Coppola_mobile,das2020non,das2019towards,das2019don,sheil2024enhancing,das2022evaluating}; sensory loss complicates use of small, low-contrast displays~\cite{Elmannai_2017_Assistive}; and limited technological proficiency further increases barriers~\cite{Finkelstein_2023_TechLiteracy}. These challenges are often more pronounced than in younger users, underscoring the need for IoT solutions thoughtfully designed to meet older adults' SPAU requirements.

Additionally, cognitive conditions such as dementia and reliance on caregivers for device operation can create avenues for exploitation in data management and access control~\cite{varghese_framework_2018,fatima2024exploring,das2020humans}. These risks demand robust, user-centered security frameworks that prevent unauthorized access and safeguard sensitive personal information. Privacy is equally critical, as IoT devices often collect and transmit health and behavioral data, raising concerns over misuse, surveillance, and loss of user control~\cite{Frik_Privacy_2019}. Yet, many interfaces overlook older adults' cognitive, sensory, and physical needs, resulting in complex navigation, small fonts, and unclear instructions. Such oversights limit IoT's potential to enhance independence and well-being, while compounding security and privacy risks. 

Prior systematization of knowledge (SoK) studies have examined SPAU challenges in IoT more broadly~\cite{varghese_framework_2018,Ashraf2022}, or focused on at-risk user groups in contexts such as healthcare technologies~\cite{wartford_framework_2021}. While these works offer valuable insights, they often treat older adults as one subset of a larger population, without tailoring criteria to their specific age-related vulnerabilities, cognitive load constraints, or accessibility needs. Our work extends this literature by introducing a SPAU framework explicitly adapted for older adults, mapping these criteria to unique threat vectors, and applying it systematically to two decades of IoT research to reveal gaps that remain unaddressed in prior reviews.

To address these interconnected challenges, we conducted a systematic review of two decades of research, screening $163$ papers and performing a framework-based evaluation of $44$ studies. Building on and extending prior SPAU frameworks, we introduce a version explicitly tailored to age-related needs, identifying critical gaps, evaluating current solutions, and analyzing threats unique to older adults' IoT use. Our work addresses the following RQs:

\begin{itemize}
    \item \textit{RQ1: Which security, privacy, accessibility, and usability (SPAU) challenges are unique to older adults in IoT adoption and use, and how do these challenges alter the threat landscape compared to the general IoT user base?}
    \item \textit{RQ2: How can existing risk assessment methodologies and threat modeling frameworks be extended to incorporate age-related cognitive, sensory, and motor limitations, and how can these extensions be operationalized within secure IoT design lifecycles?}
    \item \textit{RQ3: What user-centered design strategies, validated through empirical evaluation, most effectively deliver IoT systems that maintain robust security and privacy while ensuring accessibility and usability for older adults?}
\end{itemize}

Through this work, we make the following key~\textbf{contributions}:  
\begin{enumerate}
    \item We design the \emph{SPAU--IoT Framework}, a security-first evaluation model comprising $27$ verifiable criteria across four dimensions: security (e.g., multi-factor authentication, end-to-end encryption, secure update pipelines), privacy (e.g., local data processing, explicit opt-in consent, differential privacy techniques), accessibility (e.g., WCAG-compliant visual design, multimodal interaction support), and usability (e.g., guided task flows, adaptive error recovery). Unlike prior SPAU models, our framework explicitly encodes age-related threat exposure pathways and integrates them into the evaluation process.
    
    \item We construct an adversary-centric IoT threat model for older adults, using STRIDE to map assets, communication channels, and backend services to exploit vectors such as phishing, caregiver impersonation, credential stuffing, and insecure default configurations. Each threat is linked to attack preconditions tied to cognitive, sensory, or motor decline, and paired with concrete mitigations implementable at the device, network, or application layer.
    
    \item We perform a systematic review of $44$ peer-reviewed studies (from an initial set of $163$ identified via a reproducible multi-phase Boolean search and strict inclusion/exclusion protocol), applying the SPAU--IoT Framework to quantitatively assess the prevalence and depth of SPAU controls. Our analysis surfaces under-addressed attack surfaces, particularly at the intersection of accessibility and security, and translates these into actionable, standards-aligned recommendations for securing IoT deployments for older adults.
\end{enumerate}

\section{Background and Motivation}
IoT is increasingly embedded in daily life, transforming sectors from healthcare to home automation~\cite{9296359, 9214127,ghosh2025poster,gopavaram2021iot,streiff2019overpowered,podapati2025sok,kalhorevaluating}. Its ubiquity enables unprecedented functionality, particularly in smart homes, wearables, and connected healthcare~\cite{saka2024evaluating}, yet also introduces complex attack surfaces where sensitive personal data are collected, processed, and stored~\cite{pang2013technologies, 9296359}. Despite this potential, Sadek et al.~\cite{Sadek_2022} highlight a persistent gap in addressing the needs of underserved populations, notably older adults.

As individuals age, they often face chronic diseases, disabilities, and a greater need for assistive technologies~\cite{alkhatib_privacy_2018, garg_privacy_2014}. Cognitive, sensory, and physical declines can complicate interaction with complex interfaces~\cite{coelho2015iot, elahi_human-centered_2021, poyner_privacy_2018}, increasing vulnerability to threats such as phishing, social engineering, weak authentication, and accidental data exposure~\cite{Sahu_2021_Geriatric}. Conditions like dementia further raise the risk of unintentional data leaks, misinterpreting consent requests, or granting inappropriate device access to caregivers. Frik et al.~\cite{Frik_Privacy_2019} report that older adults often feel a loss of control over personal data when using wearables and smart home devices, fueling distrust and reduced adoption. This demographic differs from the general IoT user base in three ways: (\textit{i}) heightened cognitive and sensory vulnerabilities that alter the threat landscape, (\textit{ii}) greater dependence on others for device setup and management, and (\textit{iii}) reliance on IoT features tied to sensitive health data. These factors complicate the trade-off between autonomy and security, as noted by De Carli et al.~\cite{de_carli_vision_2021}, who advocate systems that align with older adults' capabilities while ensuring secure communications and privacy-preserving data flows. Alkhatib et al.~\cite{alkhatib_privacy_2018} similarly call for privacy models that protect independence without sacrificing safety.

Addressing these challenges requires uniting accessibility and usability with robust security and privacy controls. Coelho et al.~\cite{coelho2015iot} and Mikusz et al.~\cite{Mikusz2019SupportingOA} emphasize user-centered methodologies adapted to age-related limitations. Our work builds on this need, providing a systematic review, an evaluative \emph{SPAU--IoT Framework}, and an adversary-informed threat model to guide inclusive and secure IoT design for older adults.

\section{\lq\lq SPAU--IoT Framework\rq\rq}
The \emph{SPAU--IoT Framework} is designed to evaluate IoT systems for older adults by integrating \textbf{Security}, \textbf{Privacy}, \textbf{Accessibility}, and \textbf{Usability} criteria into a unified, demographic-aware assessment model. Unlike general IoT evaluation checklists, this framework explicitly links each criterion to \emph{age-related vulnerabilities}, including motor impairments, sensory decline, cognitive challenges, and lower technological literacy, and provides concrete evaluation procedures for system designers and auditors.

\subsection{Framework Development}
We derived the SPAU--IoT Framework through a three-step process:  
(\textit{i}) synthesizing established models in security~\cite{Bonneau_Password_2012, acharya_security_2020}, privacy~\cite{varghese_framework_2018}, usability~\cite{Ashraf2022}, and accessibility~\cite{Rus2020};  
(\textit{ii}) incorporating lessons from IoT-for-older-adults research, such as Melyani et al.'s smart house framework~\cite{Melyani_Smarthome_2018}, which lacked integration across all four SPAU dimensions; and  
(\textit{iii}) iteratively refining the criteria through expert evaluations to ensure relevance across heterogeneous IoT contexts (wearables, smart home devices, healthcare IoT). The novelty lies in \emph{systematically mapping each criterion to the security and usability threat model of older adults}, a link missing in prior work, where SPAU dimensions were considered in isolation.
We summarize how the SPAU–IoT criteria were derived from five source frameworks spanning security, privacy, usability, and accessibility. We extracted criteria from these works, normalized terminology, and mapped them to older adult vulnerabilities reported in IoT studies (e.g., memory load, motor limitations, sensory decline, caregiver mediation). Redundant or highly technical items without practical relevance for older-adult IoT contexts were merged or removed. Two researchers independently reviewed the framework across two iterations, resolving differences through discussion. ~\autoref{tab:traceability} provides a traceability summary showing representative examples of which source criteria and the rationale.

\begin{table*}[ht]
\caption{Representative traceability examples from source frameworks to SPAU--IoT criteria.}
\label{tab:traceability}
\centering
\begin{tabular}{p{3.5cm} p{4cm} p{2.5cm} p{6cm}}
\toprule
\textbf{Source Framework} & \textbf{Example Criterion} & \textbf{Action} & \textbf{Rationale} \\
\midrule
Bonneau et al. (Security) & Password memorability & Merged into S3 & Memory constraints common in older adults \\
Acharya et al. (Security) & Device \& User authentication & Kept as S1, S3 & Direct relevance to IoT threat mitigation \\
Varghese et al. (Privacy) & Information leakage & Kept as P7 & Emphasizes leakage of sensitive metadata and behavioral data \\
Ashraf et al. (Usability) & Error recovery mechanisms & Merged into U1, A5 & Combined with accessibility guidance for fault tolerance \\
Rus et al. (Accessibility) & Motor and cognition-friendly interaction design & Kept as A3 & Emphasizes tangible, easy-to-handle controls for older adults \\
\bottomrule
\end{tabular}
\end{table*}

\subsection{Security Criteria}
Older adults' interaction patterns and constraints significantly alter the IoT threat landscape: reduced motor agility, weaker short-term memory, reliance on caregivers, and lower digital literacy increase susceptibility to exploitation. To address this, the SPAU--IoT Framework defines nine security criteria:

\begin{itemize}
    \item[S1] \textit{Resilient-to-Cyber-Threats:} Systems must mitigate common attack vectors such as phishing, malware injection, and device hijacking. Older adults are disproportionately targeted by social engineering~\cite{Oliveira_2017_Dissecting}; built-in safeguards (e.g., suspicious activity detection, simplified security alerts) and in-context user education are essential.
    
    \item[S2] \textit{Resilient-to-Throttled/Unthrottled-Guessing:} Older adults often select weak, easily guessable passwords~\cite{pilar_2012_passwords, ray_2021_older}. Systems should enforce intelligent rate-limiting, account lockout, salted hashing, and credential stuffing prevention to defend against both manual and automated guessing.
    
    \item[S3] \textit{Secure Authentication Mechanisms:} Authentication must balance robustness and cognitive load. Alternatives such as biometric verification (fingerprint, facial recognition) or simplified MFA (PIN + possession factor) reduce password dependence while maintaining security.
    
    \item[S4] \textit{Data Encryption:} Apply standards-based encryption (e.g., AES-256) for both data in transit and at rest. Health and financial information, often collected passively via wearables or smart home sensor must be transmitted securely to caregivers or medical professionals.
    
    \item[S5] \textit{Secure Communication:} Enforce TLS/SSL or equivalent secure protocols without requiring manual configuration. Communication security should extend to both device–cloud and device–device channels, preventing man-in-the-middle and replay attacks.
    
    \item[S6] \textit{Secure-by-Default:} Devices should ship with hardened configurations (e.g., firewall rules, disabled unused ports) pre-enabled, eliminating the need for users to perform complex setup. This is critical for users with limited technical skills.
    
    \item[S7] \textit{Secure Software Update:} Implement automated, code-signed, encrypted updates to patch vulnerabilities promptly. Updates should require minimal or no user action to reduce exposure windows caused by delayed patching.
    
    \item[S8] \textit{Secure Incident Response:} Vendors must maintain a breach-handling protocol covering detection, containment, and recovery. Communication to older adults should use plain, non-technical language with actionable instructions, avoiding panic while ensuring swift action.
    
    \item[S9] \textit{Guardianship Features:} Provide controlled, permission-scoped access for authorized caregivers to configure settings or respond to alerts, with strict authentication to prevent abuse. 
\end{itemize}
By aligning each security criterion with both established technical best practices and the distinct usage contexts of older adults, the SPAU--IoT Framework advances beyond generic IoT security checklists to deliver a demographic-aware, operationalizable model that supports secure, accessible, and confidence-building system design.

\subsection{Privacy Criteria}
The privacy dimension of the SPAU--IoT Framework addresses the protection of sensitive personal, health, and behavioral data frequently processed by IoT systems used by older adults. Each criterion integrates established privacy engineering practices with design considerations that ensure transparency, control, and trust.

\begin{itemize}

\item[P1] \textit{Data Minimization:} Collect and retain only the minimum data required for the system's core functionality, particularly when handling health or behavioral information. Reducing the volume and duration of stored data lowers the attack surface, mitigates the impact of potential breaches, and aligns with data protection regulations.

\item[P2] \textit{User Consent and Control:} Provide clear, accessible mechanisms for users to grant, review, and revoke consent for data collection and processing. Interfaces should enable straightforward configuration of access permissions, ensuring that individuals can easily manage who can view, share, or process their data.

\item[P3] \textit{Privacy-Preserving Analytics:} Employ techniques such as data anonymization, aggregation, and differential privacy to enable analysis without exposing identifiable information. This ensures that valuable insights (e.g., for health monitoring) can be derived while maintaining confidentiality.

\item[P4] \textit{Data Retention and Disposal:} Define and enforce explicit retention policies, keeping data only as long as necessary for its stated purpose. Implement secure deletion methods (e.g., cryptographic erasure) and communicate these practices transparently to users.

\item[P5] \textit{Requiring Explicit Consent:} Design explicit consent workflows for activities involving sensitive data, avoiding ambiguous defaults or hidden opt-ins. Consent prompts should be clear, concise, and accessible, supporting informed decision-making.

\item[P6] \textit{Privacy Impact Assessment (PIA):} Conduct PIAs before deployment to identify privacy risks, evaluate compliance with regulations, and ensure safeguards are appropriate for the system's data sensitivity and user context. PIAs should include scenarios involving caregiver or third-party access.

\item[P7] \textit{Information Leak Prevention:} Integrate mechanisms to detect and prevent unauthorized data disclosure, including strict access controls, continuous monitoring, and anomaly detection for unusual access patterns. 

\item[P8] \textit{Third-Party Access Restriction:} Enforce fine-grained authorization policies that limit data sharing to explicitly approved, trusted entities. For scenarios like remote health monitoring, apply strong authentication and secure channels to ensure that only verified parties can access sensitive information.
\end{itemize}

\subsection{Accessibility Criteria}
The accessibility dimension of the framework focuses on ensuring that IoT systems are usable by individuals with diverse physical, sensory, and cognitive abilities, including those that may emerge with age. Each criterion aligns with established accessibility engineering practices and relevant legal and technical standards.

\begin{itemize}

\item[A1] \textit{Compliance with Disability Legislation:} IoT solutions should adhere to legal requirements such as the Americans with Disabilities Act (ADA) and equivalent international regulations. Compliance promotes inclusive design by incorporating features like voice-activated control, high-contrast visual modes, adjustable font sizes, tactile feedback, and compatibility with assistive devices.

\item[A2] \textit{Compatibility with Assistive Technologies:} Seamless interoperability with devices such as hearing aids, screen readers, braille displays, and mobility aids is essential. Systems should support multimodal interaction (e.g., voice, touch, gesture) and adapt to varied speech patterns, hearing ranges, and visual needs without additional configuration burdens.

\item[A3] \textit{Universal Accessibility:} Incorporate universal design principles that accommodate a wide range of physical, sensory, and cognitive capabilities. Examples include simplified navigation structures, consistent iconography, scalable user interface elements, and customizable interaction speeds. This approach minimizes adaptation costs and maximizes usability across diverse user groups.

\item[A4] \textit{Compliance with Accessibility Standards:} Align with technical standards such as the Web Content Accessibility Guidelines (WCAG), ISO $9241$ (Ergonomics of Human-System Interaction), and industry-specific protocols. Standards compliance ensures interoperability with accessibility tools and reduces barriers that can lead to indirect security or privacy risks.

\item[A5] \textit{Error Forgiveness:} Design fault-tolerant interfaces that anticipate and recover from input errors without penalizing the user. Features such as confirmation prompts for high-impact actions, undo/redo capabilities, and context-sensitive guidance in plain language can prevent accidental configuration changes and maintain user confidence.
\end{itemize}

\subsection{Usability Criteria}
The usability dimension of the framework focuses on designing systems that minimize cognitive load, reduce operational complexity, and promote user confidence over time. Each criterion draws from established HCI principles while accounting for interaction contexts common among older adult users.

\begin{itemize}
\item[U1] \textit{Guided Interaction:} Minimize setup friction and streamline reconfiguration tasks by reducing the number of steps and decision points. Employ task-oriented workflows, clear visual cues, and context-aware prompts to support error-free completion of core functions, thereby lowering cognitive effort and improving task efficiency.

\item[U2] \textit{Clear System Feedback:} Provide unambiguous, multimodal feedback that communicates system status, progress, and errors in real time. Feedback should be perceivable via multiple channels: visual (large, high-contrast text/icons), auditory (spoken cues or alerts), and, where appropriate, haptic signals, ensuring accessibility across varying sensory profiles.

\item[U3] \textit{Integrated Assistance:} Embed contextual support directly into the interface through interactive tutorials, step-by-step wizards, and searchable help modules. Intelligent assistance mechanisms, such as voice agents or conversational chatbots, should enable on-demand guidance without requiring users to navigate away from their primary task.

\item[U4] \textit{Consistent User Experience:} Maintain uniformity in visual design, terminology, navigation structure, and interaction logic across all device functions and companion applications. Consistency reduces the learning curve, fosters recognition over recall, and prevents usability regressions when switching between related features.

\item[U5] \textit{Progressive Learning Path:} Implement tiered exposure to system capabilities, starting with essential features and gradually introducing advanced functions. This staged approach allows skill acquisition at an individual's preferred pace and prevents overwhelming users during initial onboarding.
\end{itemize}

\section{Method}
\label{Sec:Method}
\begin{figure}[ht!]
    \centering
    \resizebox{0.7\columnwidth}{!}{%
    \begin{tikzpicture}[node distance=0.3cm and 0.5cm] 

        \node (identification) [label, minimum height=0.5cm] {Identification};
        \node (records) [process, below=0.2cm of identification, text width=3.3cm] 
        {Keyword Search:\\ Google Scholar (n=5200), IEEE (n=389), ACM DL (n=170), Springer (n=51),\\ Sci. Direct (n=43), PubMed (n=49)};
        \node (removed) [process, right=0.5cm of records, text width=3.3cm] 
        {Duplicates removed:\\ (n=464)};

        \node (screening) [label, below=0.2cm of records] {Screening};
        \node (screened) [process, below=0.2cm of screening, text width=3.3cm] 
        {Papers after deduplication:\\ (n=5438)};
        \node (excluded) [process, right=0.5cm of screened, text width=3.3cm] 
        {Excluded (not relevant to RQs):\\ (n=3982)};
        
        \node (retrieval) [process, below=0.2cm of screened, text width=3.3cm] 
        {Title \&\\Abstract Screening:\\ (n=1456)};
        \node (notretrieved) [process, right=0.5cm of retrieval, text width=3.3cm] 
        {Excluded:\\ (n=1293)};

        \node (eligibility) [label, below=0.2cm of retrieval] {Eligibility};
        \node (assessed) [process, below=0.2cm of eligibility, text width=3.3cm] 
        {Full-Text Screening:\\ (n=163)};
        \node (excluded_eligibility) [process, right=0.5cm of assessed, text width=3cm] 
        {Excluded:\\ (n=119)};

        \node (included) [label, below=0.2cm of assessed] {Included};
        \node (included_reports) [process, below=0.2cm of included, text width=3.3cm] 
        {Papers in final review:\\ (n=44)};

        \draw [arrow] (identification.south) -- (records.north);
        \draw [arrow] (records.east) -- (removed.west);
        \draw [arrow] (records.south) -- (screening.north);
        \draw [arrow] (screening.south) -- (screened.north);
        \draw [arrow] (screened.east) -- (excluded.west);
        \draw [arrow] (screened.south) -- (retrieval.north);
        \draw [arrow] (retrieval.east) -- (notretrieved.west);
        \draw [arrow] (retrieval.south) -- (eligibility.north);
        \draw [arrow] (eligibility.south) -- (assessed.north);
        \draw [arrow] (assessed.east) -- (excluded_eligibility.west);
        \draw [arrow] (assessed.south) -- (included.north);
        \draw [arrow] (included.south) -- (included_reports.north);

    \end{tikzpicture}
    }
    \caption{PRISMA Diagram Summarizing Study Design}
    \Description{PRISMA flowchart showing identification, screening, eligibility, and inclusion stages of paper selection.}
    \label{fig:PRISMA}
\end{figure}
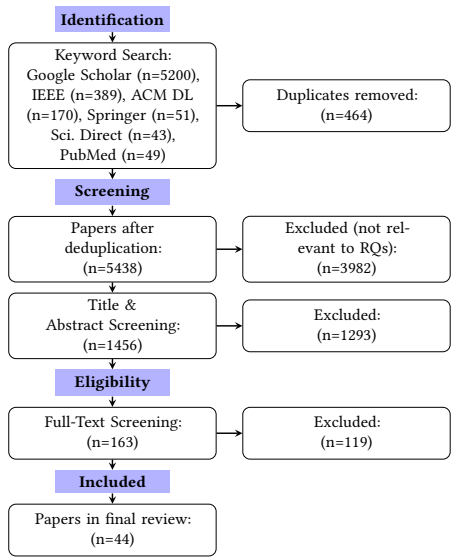

For our systematic literature review, we followed a five-step approach based on the methodologies outlined by prior works of Stowell et al. \cite{stowell_designing_2018} and Tazi et al.~\cite{tazi_sok_2021}. The steps of the review included: \textit{database search, title and abstract screening, full-text screening, data extraction, and analysis}. We chose this approach as it offers a systematic framework for conducting our analysis. Our study covers the period from $2004$ to $2024$ and specifically focuses on the security, privacy, accessibility and usability of IoT among older adults, a critical area that has received limited attention in prior research. To ensure the relevance and quality of the selected papers, we started with a planning phase where we set our \textit{inclusion criteria:} Published in peer-reviewed publications, Focused on older adults as defined in our study, Concentrated on IoT from the perspective of older adults, and Included only full papers written primarily in English. \textit{exclusion criteria:} Work-in-progress papers, extended abstracts, and non-peer-reviewed studies were excluded. 

\subsection{Search Strategy (Keywords \& Database)}
Our search keywords included terms such as ``Internet of Things," ``IoT," ``Internet of Everything," ``Smart Devices," ``Connected Devices," ``Smart Homes," ``Smart Cities," ``Wearable," ``Security," ``Privacy," ``Accessibility," ``User Studies," and ``Usability." From the user angle, we focused on the following keywords: ``Older Adult," ``Elderly," and ``Seniors~\footnote{Although the terms ``Elderly" or ``Seniors" are not considered ideal for referring to older adults~\cite{Ageinclusive}, we have still retained these keywords as they are still commonly used in existing literature.}" After we confirmed our initial set of keywords, suitable databases were identified for our search. We referred to digital databases used in some previous systematic reviews relevant to socio-technical studies~\cite{el_bekkali_systematic_2022, das2019evaluating, aqeel_review_2022, tazi_sok_2021}. The digital libraries we used included IEEE Explore, ACM Digital Library, Science Direct, Springer, PubMed, and Google Scholar. We developed an iteratively generated boolean search string across the technology and the understudied population to conclude the search strategy before commencing our search operation. Our data collection and elimination process flow system is shown in~\autoref{fig:PRISMA}.

\subsection{Keywords-based Search}
We initiated our search by employing boolean connectors and the predetermined set of keywords from the previous phase to conduct a keywords-based search in each of the selected digital libraries. Our search terms consisted of the following sets: \textit{(Internet of Things OR IoT OR Internet of Everything OR Smart Devices OR Connected Devices OR Smart Homes OR Smart Cities OR Wearable) AND (Security OR Privacy OR Accessibility OR Usability OR User Studies)} AND \textit{(Older Adult OR Elderly OR Senior)}. After searching our terms across various digital libraries, we collected a total of $47,317$ full-text papers. We then narrowed our keyword search to abstracts, resulting in a reduction to $5,902$ papers.
We explored web tools for screening systematic reviews and identified several tools, including Covidence~\cite{covidence}, Rayyan~\cite{rayyan}, EPPI-Reviewer~\cite{eppi}, and DistillerS~\cite{distillersr}. After a careful evaluation, we selected Rayyan for its open-source nature, suitability to our objectives, and ability to enhance screening effectiveness and trackability. We imported our $5,902$ papers into Rayyan, and the initial screening objective was to remove duplicate papers using the auto-deduplication function, which resulted in the removal of $464$ duplicates. 

Rayyan, a web tool, provides a user-friendly interface for efficiently screening titles and abstracts.  After eliminating the duplicates, we had $5,438$ papers for further screening. We utilized the keyword inclusion and exclusion feature provided by Rayyan to narrow the search to the title and abstract fields only, rather than the original full text explored during the initial keywords-based search. 
Due to our established criteria, we excluded $3,982$ papers that were not relevant to our research question. We were left with $1,456$ papers, which qualified for the title and abstract screening phase of the selection process. It's crucial to note that certain papers may mention the keywords in their full text, but they did not significantly address the specific topic we are investigating. 

Our Boolean query was initially 
executed at the full-text level across all databases (returning $47,317$ 
papers), then refined to title and abstract fields only to improve 
relevance (returning $5,902$ papers). These $5,902$ papers were imported into Rayyan, where we used the tool solely for: (1) automatic duplicate 
detection based on DOI/title matching ($464$ duplicates removed), and (2) 
keyword-based inclusion/exclusion filtering on titles and abstracts to 
remove clearly irrelevant papers ($3,982$ excluded). No AI-based content 
analysis or automated decision-making was employed; all screening decisions were made by human researchers. The remaining $1,456$ papers underwent independent title/abstract screening by two researchers (inter-rater agreement: $91\%$), returning $163$ papers for full-text review. After independent full-text screening by both researchers (inter-rater 
agreement: $89\%$), $44$ papers met our inclusion criteria and were included in the final analysis. A complete list of these $44$ papers is provided in Appendix.

\subsection{Title and Abstract Screening}
We considered only papers that are specifically relevant to IoT security, privacy, accessibility or usability considerations specifically for older adults. In this screening phase, we examined titles and abstracts to determine relevance to IoT security, privacy, accessibility and usability for older adults. Titles were deemed relevant if they contained relevant keywords, and abstracts were examined for indications of importance to our research idea. After we took this approach, we excluded papers from our corpus that were unrelated to our research scope. We identified a total of $163$ papers that were relevant and these were advanced to the full-text screening stage.

\subsection{Full-Text Screening}
To ensure that the data collected was of high quality and relevance to our research questions, we conducted a quality assessment of each selected article based on our inclusion and exclusion criteria. The use of methodological quality assessment was also a crucial step in our evaluation process. We used Zotero~\cite{zotero_homepage} to ensure thorough screening and organization of the collected papers. During the full-text screening phase, each article was carefully examined to determine whether it met our study's criteria. To address our research questions, relevant information from our selected papers was extracted and synthesized. The final collection of papers chosen for evaluation was the outcome of the full-text screening process.
In this stage, two researchers of this work analyzed each paper's content to ensure it aligned with our research goals. Our objective was to evaluate the security, privacy, accessibility and usability of IoT and smart devices utilized by older adults, encompassing various devices like sensors, wearable, and mobile-based technologies, to ensure the inclusion of all relevant components. The selected papers were then critically evaluated to extract the most pertinent information required for our study. The contents of the chosen papers were synthesized and analyzed in the subsequent section. After this evaluation process, we were able to select $44$ papers that met the inclusion criteria for our study from the $163$ papers that remained after our screening. The remaining $119$ papers that were eliminated addressed subjects including general ambient assistive living systems, non IoT-related technologies, and populations outside the established definition of older adults ($65$ and above).

\subsection{Data Analysis and Synthesis}
We conducted an in-depth analysis of the $44$ papers selected for this phase, starting with data collection. To capture essential information from the selected papers, such as study design, methodology, motivation, data collection method, type of IoT, and key findings related to IoT security, privacy, accessibility and usability for Older Adults, we created a data extraction form. We then proceeded to data synthesis, guided by our research questions and study objectives. During this process, we summarized key findings across studies, identified patterns, and observed any gaps or inconsistencies in the literature. Our goal was to synthesize a thorough and meaningful summary of the relevant literature, which can help identify crucial research gaps and inform the development of best practices and guidelines for promoting the security, privacy, accessibility and usability of older adults in the context of IoT. Besides summarizing the key findings across studies and identifying patterns, gaps, or inconsistencies, we also carefully evaluated the quality of the evidence presented in each article. We assessed the strength of the evidence and the validity of the study design to determine the overall confidence we could place in the findings. This assessment allowed us to make informed judgments about the implications of the literature for the development of policies, interventions, or other strategies to enhance the security, privacy, accessibility and usability of older adults in the context of IoT. Ultimately, our review is expected to provide essential insights into the challenges and opportunities associated with IoT security, privacy, accessibility and usability for older adults and can help guide future research and practice in this domain.

We organized our extracted data using a codebook-driven thematic analysis, allowing us to identify commonalities and distinctions across the reviewed studies. This process involved mapping each paper to specific aspects of IoT security, privacy, accessibility, and usability (SPAU) relevant to older adults. From this synthesis, three dominant thematic clusters emerged: \textit{Healthcare and Wellness Applications}, \textit{Independent Living Solutions}, and \textit{Legislation \& Policies}. While each theme is grounded in the lived contexts of older adults, our analysis also incorporated the underlying technical considerations, which are discussed in the subsequent section alongside our proposed threat model.

\begin{itemize}
    \item \textit{Healthcare and Wellness Applications:}  
    Forty-two studies examined IoT adoption for health-related use cases, including remote patient monitoring, chronic disease management, early intervention, and wellness promotion. Examples include wearable fall-detection systems~\cite{ganesh_iot-based_2019, Chavan_2017, Fan_2022}, daily routine guidance platforms~\cite{suzuki_updated_2018}, and vital sign tracking for high-risk patients~\cite{philip_internet_2021}. Other works explored passive monitoring of mobility or behavioral anomalies to flag potential risks. 
    
    \item \textit{Independent Living Solutions:}  
    Several studies investigated IoT integration in home and assisted living environments~\cite{Jo_Elderly_2021, Geng_2022, Eklund2005, Visutsak2017}. Applications included hazard detection, mobility assistance, daily activity support, and nutrition tracking. For instance, Fang et al. proposed a non-intrusive sensor network to monitor movement within the home~\cite{fang_nonintrusive_2021}. However, balancing supportive automation with user autonomy emerged as a recurring concern. 
    
    \item \textit{Legislation \& Policies:}  
    Studies highlighted the importance of collaboration between policymakers and designers to establish standards that safeguard personal data while ensuring accessibility and usability~\cite{Vitak_2020, KnightThora2022}. Debate persists over whether such standards should be mandatory or voluntary, with arguments centering on incentivizing compliance without unduly restricting user choice~\cite{davis2014bridging}. 
\end{itemize}

\section{Results and Discussion}
\label{Sec:Results}
\subsection{SPAU Challenges (RQ1)}
\subsubsection{\textbf{IoT Adoption and Use Among Older Adults}}
Our study demonstrated that the primary application of IoT adoption centers around smart home solutions for older adults, with over $76\%$ of the papers exploring this area. The emphasis on smart homes is driven by the increasing population of older adults and the ongoing need for sustainable assisted living solutions, particularly in the medical and health domain~\cite{alkhatib_privacy_2018}. Consequently, a significant portion of the research focuses on addressing medical or health-related concerns. Throughout our data extraction process, we observed that at least $91\%$ of papers in our corpus discussed IoT as a monitoring instrument. This finding suggests that the majority of IoT adoption for older adults aims to keep a close eye on their well-being. However, there is potential to expand the scope of IoT applications for older adults beyond monitoring. Some papers already combine monitoring with assistance, navigation, detection, and support as shown in~\autoref{tab:study_type}. By integrating these multiple dimensions, researchers aim to improve the quality of life and safety of older adults while catering to their unique needs and challenges.

To further enhance the impact of IoT on older adults' lives, researchers could explore other aspects such as social engagement, mental health, and personalized care. For instance, IoT could facilitate communication and connection with friends, family, and caregivers, thereby addressing loneliness and social isolation~\cite{Mendel_relative_2020}. Moreover, IoT applications could be developed to monitor cognitive health by screening for early signs of cognitive decline and providing personalized interventions to support older adults' mental well-being. Several papers in our corpus emphasized various monitoring techniques employed by IoT for older adults making it essential to preserve their security and privacy. Besides the need for assistive technologies, we noticed that IoT sensors play a crucial role in the monitoring process. Therefore, our study identified some monitoring techniques which we have divided into Physiological, Functional and Physical monitoring as mentioned in this paper.

\begin{enumerate}
\item \textbf{Physiological Monitoring:}
Some papers~\cite{poyner_privacy_2018, mulero_iot-aware_2018, dobre_improving_2019, garg_privacy_2014, zavalyshyn_smart_2021, philip_internet_2021, khan_iot-based_2019, guizani_iot_2020, paraschiv_iot_2021, pinto_we-care_2017, suzuki_updated_2018} reported that physiological monitoring involves tracking vital signs, such as body temperature, blood pressure, and pulse rate, in real-time for older adults. Wearable sensors often provide accurate readings and must be non-invasive to avoid user discomfort. This type of monitoring reduces risks significantly, proving particularly beneficial for older adults with chronic health concerns. Without such intervention, the fatality rate among most older adults may increase exponentially. 
Additionally, physiological monitoring through IoT enables timely intervention and therapy, allowing for early detection of health issues. However, the security and privacy of IoT for older adults must be carefully addressed to ensure the confidentiality and privacy of their health information.

\item \textbf{Functional Monitoring:}
This type of monitoring evaluates older adults' ability to perform daily activities that includes walking, bathing, dressing, and eating, as highlighted in some papers~\cite{poyner_privacy_2018, mulero_iot-aware_2018, dobre_improving_2019, garg_privacy_2014, zavalyshyn_smart_2021, philip_internet_2021, khan_iot-based_2019, paraschiv_iot_2021, ganesh_iot-based_2019, fang_nonintrusive_2021, almeida_performance_2018, pinto_we-care_2017, suzuki_updated_2018}. This monitoring type aids in identifying any decline in older adults' capacity to maintain their usual routine and must occur in real-time to ensure prompt intervention. Security and privacy surrounding such IoT are vital because breaches could result in transmitting incorrect information, jeopardizing older adults' health. Our review did not encounter any threat analysis among the papers, so we could not assess this intervention's threats.

\item \textbf{Physical Monitoring:}
This focuses on environmental hazards, such as gas leaks, fire outbreaks, and human threats, such as intruders. Older adults benefit from a safety system that alerts them to these dangers. Some papers~\cite{mulero_iot-aware_2018, elkahlout_iot-based_2020, garg_privacy_2014, philip_internet_2021, khan_iot-based_2019, elahi_human-centered_2021, guizani_iot_2020, pinto_we-care_2017, suzuki_updated_2018} discuss specific IoT designed for this purpose, which support all other security, privacy, accessibility and usability for older adults. Papers on this monitoring type underscore the use of environmental sensors and actuators to detect any leakage or hazards harmful to older adults. Notably, Fang et al.~\cite{fang_nonintrusive_2021} proposed a non-intrusive pyroelectric infrared (PIR) sensing device for smart home monitoring of living-alone older adults to track their movement.
While we recognize that monitoring devices designed for older adults inherently consider usability and comfort during design, our analysis reveals that ensuring underlying data security remains a persistent challenge. The transmission mechanisms between wearable sensors, processing hubs, and cloud storage often lack standardized encryption, creating vulnerabilities for attacks. Similarly, access control policies for IoT show inconsistency across monitoring system implementations. Our focus is not on suggesting deficiencies in the device interface for older users. Instead, we emphasize the need to strengthen the security protections for user data that these monitoring solutions handle. 
\end{enumerate}

\subsubsection{\textbf{Existing Approaches and Study Type}}
All the papers in our corpus are categorized in table~\ref{tab:study_type}, and the analysis of these papers was explored further in the paper. By examining the final dataset, we identified comparison issues which allowed us to compare different IoT security, privacy, accessibility or usability approaches in those studies involving older adults. We considered factors such as the types of IoT deployed accessibility or usability mechanisms in place, user experience, and efficiency of security and privacy measures. By adopting the Stowell and Faiza technique, we identified patterns and trends in the data, providing us with an overview of the current state of IoT security, privacy, accessibility and usability for the older population. We found that some motivations for using IoT in this context include special medical care~\cite{elkahlout_iot-based_2020, dobre_improving_2019, garg_privacy_2014, philip_internet_2021,khan_iot-based_2019, guizani_iot_2020, paraschiv_iot_2021, ganesh_iot-based_2019, fang_nonintrusive_2021, almeida_performance_2018, pinto_we-care_2017} for older adults, improved living conditions~\cite{mulero_iot-aware_2018, elahi_human-centered_2021, KnightThora2022}, suitable home automation~\cite{zavalyshyn_smart_2021, perez_review_2023, suzuki_updated_2018, pal_smarthome2017}, and compliance with relevant legislation and regulations~\cite{poyner_privacy_2018, maswadi_systematic_2020} concerning data security and privacy. In addition, the purpose of Frik et al.~\cite{Frik_Privacy_2019} was to identify older adults' attitudes toward security and privacy. We explored the type of study carried out in all the papers, which can be summarized into \textit{Technical Analysis, Prototype, and User Study} as shown in~\autoref{tab:study_type}.
 
\paragraph{Technical Analysis:}
These studies focused on technical aspects of IoT for older adults such performance evaluations, design analyses, and examinations of specific dimensions such as security, privacy, accessibility, and usability. Poyner et al. conducted a review on IoT security and privacy policies in the healthcare domain, although their analysis fell short of addressing the specific concerns and challenges faced by older adults within the studied healthcare frameworks~\cite{poyner_privacy_2018}. Almeida et al. focused their efforts on evaluating the performance of IoT monitoring systems, with a particular emphasis on ensuring the safety and well-being of older adult users~\cite{almeida_performance_2018}. De Carli et al. identified three critical areas : autonomy, control, and delegation, that should be the focal points when designing IoT network security systems tailored for older adults~\cite{de_carli_vision_2021}. Their work highlights the importance of empowering this demographic while also protecting their privacy and ensuring their security within the IoT ecosystem. 

Several studies within our dataset explored the accessibility and usability dimensions of IoT for older adults. Yun et al. and Ashraf et al. examined frameworks for evaluating usability and design considerations of IoT applications tailored to the needs of this demographic~\cite{Yun2016, Ashraf2022}. However, these works did not investigate the security and privacy implications of their proposed solutions. Similarly, Visutsak et al. and Lin et al. looked into the user interface design and cognitive accessibility for IoT intended for older adults~\cite{Visutsak2017, Lin2020}. While their contributions shed light on these crucial aspects, they did not provide a thorough analysis of the potential security and privacy risks associated with the adoption. Notably, none of the existing studies presented a complete framework that integrated all four dimensions – security, privacy, accessibility, and usability – into the design and assessment of IoT for older adults. Our work addresses this critical gap by proposing an approach that considers the specific needs and preferences of this population throughout the development lifecycle of IoT.

\begin{table*}[hbt!]
\caption{Assessment of Collected Papers Based on SPAU-IoT Framework. Here SPAU: Security, Privacy, Accessibility and Usability.  Fully Discussed Criteria = \yes | Partially Discussed = $\circ$  | Not Discussed = \no }
  \centering
  \setlength{\tabcolsep}{2.5pt}
      \footnotesize
\begin{tabular}{     
                c      
                *{9}{c}
                *{8}{c}
                *{5}{c}
                *{5}{c}
                *{4}{c}     
                *{3}{c}     
               }

               &

\multicolumn{9}{c}{{\textbf{Security}}} &
\multicolumn{8}{c}{{\textbf{Privacy}}} &
\multicolumn{5}{c}{{\textbf{Accessibility}}} &
\multicolumn{5}{c}{{\textbf{Usability}}} &
\multicolumn{4}{c}{{\textbf{IoT Application}}} &
\multicolumn{3}{c}{{\textbf{Environment}}} \\
    \textbf{Ref}  &  
    \verttext{S1}&
    \verttext{S2}&
     \verttext{S3}&
      \verttext{S4}&
       \verttext{S5}&
        \verttext{S6}&
         \verttext{S7}&
          \verttext{S8}&
           \verttext{S9}&

    \verttext{P1}&
    \verttext{P2}&
     \verttext{P3}&
      \verttext{P4}&
       \verttext{P5}&
        \verttext{P6}&
         \verttext{P7}&
          \verttext{P8}&

    \verttext{A1}&
    \verttext{A2}&
     \verttext{A3}&
      \verttext{A4}&
       \verttext{A5}&

    \verttext{U1}&
    \verttext{U2}&
     \verttext{U3}&
      \verttext{U4}&
       \verttext{U5}&
    
     \verttext{Monitoring}&
    \verttext{Navigation} &
    \verttext{Care \& Support} &
    \verttext{Detection} &
    
    
    \verttext{Smart Homes} &
    \verttext{Smart Cities} &
    \verttext{Healthcare}   \\
   
    \hline
    %
    %
\rowcolor{gray!50}
~\cite{alkhatib_privacy_2018}  &  
 \leftrule{\no}& \no& \no& \no&\no &\no &\no &\no &\no &
 \leftrule{$\circ$}&\yes &\no &\no &$\circ$ &\no & $\circ$ &\yes &
 \leftrule{\no} &\no & \no& \no&\no & 
 \leftrule{\no} &\no &\no &\no &\no & 
 
    \leftrule{\yes}&\no & \no & \no &
    \leftrule{\no}&\no&\yes \\

    ~\cite{almeida_performance_2018} &  
\leftrule{$\circ$} & \no &$\circ$ & \yes& \yes&\no &\no &\no &\no &
  \leftrule{\no}&\no & \no&\no &\no &\no & \no&\no &
 \leftrule{\no} &\no & \no& \no&\no & 
 \leftrule{\no} &\no &\no &\no &\no & 
      \leftrule{\yes}&\no&\no & \no &
    \leftrule{\yes}&\yes&\no \\

    \rowcolor{gray!50}
~\cite{elkahlout_iot-based_2020} & 
\leftrule{$\circ$} & \no &$\circ$ & \yes& \yes&\no &\no &\no &\no &
  \leftrule{\no}&\no & \no& \no&\no &\no & \no&\no &
 \leftrule{\no} &\no & \no& \no&\no & 
 \leftrule{\no} &\no &\no &\no &\no & 
 
 \leftrule{\yes}&\no&\no & \no &
    \leftrule{\no}&\yes&\no \\
~\cite{garg_privacy_2014}   &

\leftrule{\no} &\no &\no &\no &\no &\no &\no & \no&\no &
 \leftrule{\yes}&\yes &$\circ$ &\no &\yes &\no & $\circ$ &\yes &
 \leftrule{\no} &\no & \no&\no &\no & 
 \leftrule{\no} &\no &\no &\no &\no & 
 
    \leftrule{\yes}&\yes&\yes & \yes &
    \leftrule{\yes}&\no&\yes \\

\rowcolor{gray!50}
~\cite{gochoo_towards_2021}& 
 \leftrule{\no} &\no & \no& \yes&\yes &\no & \no&\no &\no &
  \leftrule{\no}&\no & \no&\no &\no &\no & \yes&\no &
 \leftrule{\no} &\no & \no&\no &\no & 
 \leftrule{\no} &\no &\no &\no &\no & 
 
 \leftrule{\yes}&\no&\yes & \no &
    \leftrule{\yes}&\no&\no \\  
    
 ~\cite{guizani_iot_2020} &
\leftrule{\yes} &\no &\yes &\yes &\yes &\no &\no & \no&\no &
  \leftrule{\no}&\no & \no&\no &\no &\no & \no&\no &
 \leftrule{\no} &\no & \no&\no &\no & 
 \leftrule{\no} &\no &\no &\no &\no & 
   \leftrule{\yes}&\no&\no & \no &
    \leftrule{\no}&\no&\yes \\

\rowcolor{gray!50}
     ~\cite{Vitak_2020} & 
\leftrule{\yes} &\no &\yes &\yes &\yes &\no &\no & \no&\no &
  \leftrule{\yes}&\yes & \no&\yes &\yes &\no & \yes&\yes &
 \leftrule{\no} &\no & \no&\no &\no & 
 \leftrule{\no} &\no &\no &\no &\no & 
       \leftrule{\yes}&\no&\yes & \no &
    \leftrule{\no}&\no&\yes \\

   ~\cite{KnightThora2022} & 
\leftrule{\yes} &\no &\yes &\yes &\yes &\no &\no & \no&\no &
  \leftrule{\yes}&\yes & \no&\no &\yes &\no & \yes&\no &
 \leftrule{\no} &\no & \no&\no &\no & 
 \leftrule{\no} &\no &\no &\no &\no & 
     \leftrule{\yes}&\no&\yes & \no &
    \leftrule{\yes}&\no&\yes \\

\rowcolor{gray!50}     
   ~\cite{perez_review_2023} &  
\leftrule{\yes} &\no &\yes &\yes &\yes &\no &\no & \no&\no &
  \leftrule{\yes}&\yes & \no&\yes &\yes &\no & \yes&\no &
 \leftrule{\no} &\no & \no&\no &\no & 
 \leftrule{\no} &\no &\no &\no &\no & 
     
     \leftrule{\no}&\no&\yes & \no &
    \leftrule{\yes}&\no&\yes \\
     ~\cite{philip_internet_2021} & 
\leftrule{\yes} &\no &\yes &\yes &\yes &\yes &\no & \no&\no &
  \leftrule{\yes}&\yes & \no&\yes &\yes &\no & \yes&$\circ$ &
 \leftrule{\no} &\no & \no&\no &\no & 
 \leftrule{\no} &\no &\no &\no &\no & 
       \leftrule{\yes}&\no&\no & \no &
    \leftrule{\yes}&\no&\yes \\

\rowcolor{gray!50}  
   ~\cite{poyner_privacy_2018} & 
\leftrule{\yes} &\no &$\circ$ &\yes &\yes &\yes &$\circ$& $\circ$&\no &
  \leftrule{$\circ$}&\yes & \no&\yes &\yes &\no & \no&\no &
 \leftrule{\no} &\no & \no&\no &\no & 
 \leftrule{\no} &\no &\no &\no &\no &  
     \leftrule{\yes}&\no&\yes & \no &
    \leftrule{\no}&\no&\yes \\

    ~\cite{townsend_privacy_2011} & 
\leftrule{\yes} &\no &\yes &\yes &\yes &\no &\no & \no&\no &
  \leftrule{$\circ$}&\yes & \no&\yes &\no &\no & \no&\no &
 \leftrule{\no} &\no & \no&\no &\no & 
 \leftrule{\no} &\no &\no &\no &\no & 
      
      \leftrule{\yes}&\no&\yes & \no &
    \leftrule{\yes}&\no&\no \\

\rowcolor{gray!50}
~\cite{zavalyshyn_smart_2021} &
\leftrule{$\circ$} &\no &$\circ$ &\yes &$\circ$ &\no &\no & \no&$\circ$ &
  \leftrule{\yes}&$\circ$ & $\circ$ &\no &\yes &\no & $\circ$&$\circ$&
 \leftrule{\no} &\no & \no&\no &\no & 
 \leftrule{\no} &\no &\no &\no &\no & 
       
       \leftrule{\no}&\no&\yes & \no &
    \leftrule{\yes}&\no&\no \\

   ~\cite{de_carli_vision_2021} &
     
\leftrule{\yes} &\no &\yes &\yes &\yes &\yes &$\circ$ & \no&$\circ$ &
  \leftrule{\no}&\yes & \no&$\circ$ &\yes &\no & \no&\no &
 \leftrule{\no} &\no & \no&\no &\no & 
 \leftrule{\no} &\no &\no &\no &\no & 
     \leftrule{\yes}&\no&\no & \no &
    \leftrule{\yes}&\no&\no \\

 \rowcolor{gray!50}
    ~\cite{Mikusz2019SupportingOA} & 
      
\leftrule{\no} &\no &\no &$\circ$ &\no &\no &\no & \no&\no &
  \leftrule{$\circ$}&\yes & \yes&\no &\yes &\no & $\circ$ & $\circ$ &
 \leftrule{\no} &\no & \no&\no &\no & 
 \leftrule{\no} &\no &\no &\no &\no &

      \leftrule{\yes}&\no&\yes & \no &
    \leftrule{\yes}&\no&\yes \\

    ~\cite{Bhiwoo2023} & 
      
\leftrule{\yes} &\no &$\circ$ &\yes &\yes &\yes &$\circ$ & $\circ$ &\no &
  \leftrule{\no}&\no & \no&\no &\no &\no & \no&\no &
 \leftrule{\no} &\no & \no&$\circ$ &\no & 
 \leftrule{\yes} &$\circ$ &\yes &\no &\no & 
      \leftrule{\yes}&\no&\yes & \no &
    \leftrule{\yes}&\no&\no \\     

\rowcolor{gray!50}
    ~\cite{Visutsak2017} & 
      
 \leftrule{\no}& \no& \no& \no&\no &\no &\no &\no &\no &
 \leftrule{\no}&\no &\no &\no &\no &\no & \no &\no &
 \leftrule{\no} &\yes & $\circ$& \no&\no & 
 \leftrule{\no} &\no &\no &\no &\no & 
      \leftrule{\yes}&\no&\yes & \yes &
    \leftrule{\yes}&\no&\no \\     
    
    ~\cite{Yun2016} &
 \leftrule{\no}& \no& \no& \no&\no &\no &\no &\no &\no &
  \leftrule{\no}&\no &\no &\no &\no &\no & \no &\no &
 \leftrule{\no} &\no & \no& \no&\no & 
 \leftrule{\yes} &\yes &\yes &\yes &$\circ$ & 
      \leftrule{\yes}&\no&\yes & \yes &
    \leftrule{\yes}&\no&\yes \\   

\rowcolor{gray!50}
        ~\cite{Lin2020} & 
          
 \leftrule{\no}& \no& \no& \no&\no &\no &\no &\no &\no &
 \leftrule{\no}&\no &\no &\no &\no &\no & \no &\no &
 \leftrule{\no} &\no & \yes& \no&\no & 
 \leftrule{\no} &\no &\no &\no &\no & 
          
          \leftrule{\yes}&\no&\no & \yes &
    \leftrule{\yes}&\no&\yes \\

        ~\cite{Ashraf2022} &
 \leftrule{\no}& \no& \no& \no&\no &\no &\no &\no &\no &
  \leftrule{\no}&\no &\no &\no &\no &\no & \no &\no &
 \leftrule{\no} &\no & \no& \no&\no & 
 \leftrule{\yes} &\yes &\yes &\yes &$\circ$ & 
          
          \leftrule{\yes}&\no&\no & \no &
    \leftrule{\yes}&\no&\no \\

\rowcolor{gray!50}
        ~\cite{Maresova2020} &
          
 \leftrule{\yes}& \no& \yes& \yes&\yes &$\circ$ &$\circ$ &$\circ$ &\yes &
  \leftrule{$\circ$}&\no &\no &$\circ$ &\no &\no & \no &\no &
 \leftrule{\no} &\no & \no& \no&\no & 
 \leftrule{\yes} &\yes &\yes &\yes &$\circ$ & 
          \leftrule{\yes}&\no&\yes & \no &
    \leftrule{\yes}&\no&\yes \\

        ~\cite{Ghosh2023} & 
          
 \leftrule{\no}& \no& \no& \no&\no &\no &\no &\no &\no &
 \leftrule{\no}&\no &\no &\no &\no &\no & \no &\no &
 \leftrule{\no} &\no & \yes& \no&\no & 
 \leftrule{\no} &$\circ$ &\yes &\no &\no &

          \leftrule{\yes}&\no&\yes & \yes &
    \leftrule{\no}&\no&\yes \\

\rowcolor{gray!50}
            ~\cite{Tan2020} & 
              
 \leftrule{\no}& \no& \no& \no&\no &\no &\no &\no &\no &
 \leftrule{\no}&\no &\no &\no &\no &\no & \no &\no &
 \leftrule{\no} &\no & \no& \no&\no & 
 \leftrule{\yes} &\yes &\yes &\yes &$\circ$ &          
              \leftrule{\yes}&\no&\yes & \no &
    \leftrule{\no}&\yes&\no \\

          ~\cite{Daurte2018} & 
            
 \leftrule{\no}& \no& \no& \no&\no &\no &\no &\no &\no &
 \leftrule{\no}&\no &\no &\no &\no &\no & \no &\no &
 \leftrule{\no} &\no & \no& \no&\no & 
 \leftrule{$\circ$} &\no &\yes &\no &\no &      
            
            \leftrule{\yes}&\no&\yes & \no &
    \leftrule{\no}&\no &\yes \\

\rowcolor{gray!50}
           ~\cite{pal_smarthome2017} & 
 \leftrule{\no}& \no& \no& \no&\no &\no &\no &\no &\no &
 \leftrule{$\circ$}&\yes &\no &\no &\no &$\circ$ & \no &\no &
 \leftrule{\no} &\no & \no& \no&\no & 
 \leftrule{\no} &\no &\no &\no &\no &

             \leftrule{\yes}&\no&\no & \no &
    \leftrule{\yes}&\no &\no \\

          ~\cite{Alagar2019} &
 \leftrule{\no}& \no& \yes& \yes &\yes &\no &\no &\no &$\circ$ &
 \leftrule{\yes}&\no &\no &\no &\no &\no & \yes &\no &
 \leftrule{\no} &\no & \no& \no&\no & 
 \leftrule{\no} &\no &\no &\no &\no &          
            
            \leftrule{\yes}&\no&\yes & \no &
    \leftrule{\no}&\no &\yes \\

\rowcolor{gray!50}
           ~\cite{Turjamaa2019} &
 \leftrule{ \no}& \no& \no& \no& \no&\no &\no &\no &\no &
\leftrule{\no}&\no &\no &\no &\no &\no & \no &\no &
 \leftrule{\no} &\yes &\yes & \no&\no & 
 \leftrule{\yes} &\yes &\yes &\yes &\no &       
             
             \leftrule{\yes}&\no&\yes & \no &
    \leftrule{\yes}&\no &\no \\

           ~\cite{KUMAR2023110720} &

 \leftrule{ \no}& \no& \no& \no& \no&\no &\no &\no &\no &
\leftrule{\no}&\no &\no &\no &\no &\no & \no &\no &
 \leftrule{\no} &\yes &\yes & \no&\no & 
 \leftrule{\yes} &\yes &\yes &\yes &\no &    
             
             \leftrule{\yes}&\no&\yes & \no &
    \leftrule{\yes}&\no &\yes \\

\rowcolor{gray!50}
           ~\cite{Kouris2020} & 
             
 \leftrule{\yes} &\yes & \yes& \yes&\yes &\no & \no&\no &\no &
  \leftrule{\yes}&\no & \no&\no &\no &\no & \yes&\no &
 \leftrule{\no} &\yes & \yes&\no &\no & 
 \leftrule{\no} &\no &\no &\no &\no & 
       
             \leftrule{\yes}&\yes&\yes & \yes &
    \leftrule{\yes}&\no &\yes \\

           ~\cite{Fournier2020DesigningDT} &
             
 \leftrule{\yes} &\yes & \yes& \yes&\yes &\no & \no&\no &\no &
  \leftrule{\yes}&\no & \no&\no &\no &\no & \no&\no &
 \leftrule{\no} &\yes & \no&\no &\no & 
 \leftrule{\no} &\no &\no &\no &\no &    
             
             \leftrule{\yes}&\no&\yes & \yes &
    \leftrule{\yes}&\no &\yes \\

\rowcolor{gray!50}
       
   ~\cite{dobre_improving_2019} & 
     
 \leftrule{\no} &\no & \yes& \yes&\no &\no & \no&\no &\no &
  \leftrule{\no}&\no & \no&\no &\no &\no & \no&\no &
 \leftrule{\no} &\no & \no& \no&\no & 
 \leftrule{\no} &\no &\no &\no &\no & 
     \leftrule{\yes}&\no&\yes & \no &
    \leftrule{\yes}&\no&\yes \\

   ~\cite{fang_nonintrusive_2021}  & 
 \leftrule{ \no}& \no& \no& \no& \no&\no &\no &\no &\no &
  \leftrule{\yes}&\yes & \yes &\no &\no &\no & \yes&\yes &
 \leftrule{\no} &\no & \no& \no&\no & 
 \leftrule{\no} &\no &\no &\no &\no & 
     \leftrule{\yes}&\no&\no & \no &
    \leftrule{\yes}&\no&\no \\

 \rowcolor{gray!50}   
       ~\cite{ganesh_iot-based_2019} &
         
 \leftrule{ \yes}& \no& \yes& \yes& \yes&\no &\no &\no &\no &
  \leftrule{\yes}&\no & \no &\no &\no &\no & \no&\no &
 \leftrule{\no} &\no & \no& \no&\no & 
 \leftrule{\yes} &\yes &\yes &\yes &\no & 
         \leftrule{\yes}&\yes&\yes & \yes &
    \leftrule{\yes}&\no&\yes \\

   ~\cite{mulero_iot-aware_2018} & 
     
 \leftrule{ \yes}& \no& \yes& \yes& \yes&\no &\no &\no &\no &
  \leftrule{\yes}&\yes & \no &\no &\no &\no & \yes&\no &
 \leftrule{\no} &\no & \no& \no&\no & 
 \leftrule{\no} &\no &\no &\no &\no & 
     \leftrule{\yes}&\no&\yes & \no &
    \leftrule{\no}&\yes&\no \\

      \rowcolor{gray!50}
    ~\cite{paraschiv_iot_2021} & 
 \leftrule{ \yes}& \no& \yes& \yes& \yes&\no &\no &\no &\no &
  \leftrule{\yes}&\ & \no &\no &\no &\no & \yes&\no &
 \leftrule{\no} &\no & \no& \no&\no & 
 \leftrule{\no} &\no &\no &\no &\no &

      \leftrule{\yes}&\no&\yes & \no &
    \leftrule{\yes}&\no&\no \\

  ~\cite{pinto_we-care_2017} &
    
 \leftrule{ \yes}& \no& \yes& \yes& \yes&\no &\no &\no &\no &
  \leftrule{\yes}&\yes & \no &\no &\no &\no & \no&\no &
 \leftrule{\no} &\no & \no& \no&\no & 
 \leftrule{\no} &\no &\no &\no &\no & 
    
    \leftrule{\yes}&\no&\yes & \yes &
    \leftrule{\yes}&\no&\yes \\

\rowcolor{gray!50}      
    ~\cite{suzuki_updated_2018} & 
 \leftrule{ \no}& \no& \no& \no& \no&\no &\no &\no &\no &
  \leftrule{\yes}&\no & \no &\no &\no &\no & \yes&\yes &
 \leftrule{\no} &\no & \no& \no&\no & 
 \leftrule{\no} &\no &\no &\no &\no & 
 
      \leftrule{\yes}&\no&\yes & \no &
    \leftrule{\yes}&\no&\no \\

          ~\cite{Angkul2020} & 
 \leftrule{ \no}& \no& \no& \no& \no&\no &\no &\no &\no &
  \leftrule{\no}&\no & \no &\no &\no &\no & \no&\no &
 \leftrule{\no} &\yes & \no& \yes&\no & 
 \leftrule{\yes} &\yes &\yes &\yes &\no & 
            
            \leftrule{\yes}&\no&\yes & \no &
    \leftrule{\yes}&\no &\no \\

\rowcolor{gray!50}                
              ~\cite{Lee2022} & 
 \leftrule{ \yes}& \no& \yes& \yes& \yes&\yes &\no &\no &\no &
  \leftrule{\yes}&\yes & \no &\no &\no &\yes & \yes&\no &
 \leftrule{\no} &\no & \no& \no&\no & 
 \leftrule{\no} &\no &\no &\no &\no &

                \leftrule{\yes}&\no&\yes & \no &
    \leftrule{\yes}&\yes &\no \\

  ~\cite{Rus2020} & 
    
 \leftrule{ \no}& \no& \no& \no& \no&\no &\no &\no &\no &
  \leftrule{\no}&\no & \no &\no &\no &\no & \no&\no &
 \leftrule{\yes} &\yes & \yes& \yes&\no & 
 \leftrule{\yes} &\yes &\yes &\yes &\no & 
    
    \leftrule{\yes}&\no&\yes & \no &
    \leftrule{\yes}&\no &\no \\

\rowcolor{gray!50}

  ~\cite{elahi_human-centered_2021} & 
    
 \leftrule{ \no}& \no& \no& \no& \no&\no &\no &\no &\no &
  \leftrule{\yes}&\no & \yes &\no &\no &\no & \no&\no &
 \leftrule{\no} &\no & \no& \no&\no & 
 \leftrule{\no} &\no &\no &\no &\no & 
    
    \leftrule{\yes}&\yes&\yes & \yes &
    \leftrule{\yes}&\yes&\yes \\

   ~\cite{Frik_Privacy_2019} & 
     
 \leftrule{ \no}& \no& \no& \no& \no&\no &\no &\no &\no &
  \leftrule{\no}& $\circ$ & \no &\no &\no &\no & \no&\yes &
 \leftrule{\no} &\no & \no& \no&\no & 
 \leftrule{\no} &\no &\no &\no &\no &
     
     \leftrule{\yes}&\yes&\yes & \no &
    \leftrule{\yes}&\yes&\yes \\ 

\rowcolor{gray!50}
       ~\cite{pal2019} & 
 \leftrule{ \no}& \no& \no& \no& \no&\no &\no &\no &\no &
  \leftrule{\yes}& $\circ$ & $\circ$ &\yes &\yes &\no & \yes&\yes &
 \leftrule{\no} &\no & \no& \no&\no & 
 \leftrule{\no} &\no &\no &\no &\no &

         \leftrule{\yes}&\no&\no & \no &
    \leftrule{\yes}&\no&\no \\  

    \hline
    
\end{tabular}
 
    \label{tab:study_type}
\end{table*}

\paragraph{Prototype:}
Ten papers in our corpus designed prototype IoT and evaluated their associated functions to determine their usefulness and security impact on older adults. This papers~\cite {mulero_iot-aware_2018,dobre_improving_2019,ganesh_iot-based_2019,fang_nonintrusive_2021,pinto_we-care_2017, suzuki_updated_2018, paraschiv_iot_2021} developed prototypes as IoT monitoring devices tailored for use in smart environments by older adults, emphasizing both the functionality and security of the devices. Two of these papers~\cite{dobre_improving_2019, paraschiv_iot_2021} proposed their own designed services to accommodate data from connected sensors, making their services scalable. In comparison, one suggested an integrated platform for non-intrusive monitoring and support for older adults, while the other proposed it cloud services can integrate IoT to create a secure smart healthcare system. Similar to Paraschiv's IaaS, Mulero et al.\cite{mulero_iot-aware_2018} propose IoT infrastructure in smart cities that leverage middleware to produce data, improving older adults' lives through research. All prototype designs used sensors, which are ordinarily targets of attacks. Suzuki et al. created a watch-over system using a Raspberry Pi IoT combined with motion sensors in a smart home to monitor older adults continuously~\cite{suzuki_updated_2018}. They opt for sensors over conventional cameras to prioritize older adults' privacy. 

Aiming to provide better living conditions for older adults, Pinto's team designed a usable We-Care system in the form of a wristband to monitor and record older adults' vital signs~\cite{pinto_we-care_2017}. To protect wireless communication within the system, they implement linked-layer encryption. Similarly, Ganesh et al. used a Raspberry Pi in combination with sensors to design an accessible intelligent home primarily controlled by voice using Google Duplex Artificial Intelligence~\cite{ganesh_iot-based_2019}. They also created a smart bed to help monitor older adults' sleep and provide them with fall protection. However, there was often a lack of thorough incorporation of accessibility and usability considerations alongside security and privacy. This absence emphasizes a notable gap in prototype development, emphasizing the necessity for a more tailored IoT design that includes all four crucial dimensions.

\subsubsection{\textbf{Security and Privacy Risks Associated:}}
IoT has introduced numerous security and privacy hazards for older adults. Commercially available smart devices often suffer from poor security and privacy design, leaving them vulnerable to exploitation by attackers~\cite{Morrison_older_2021}. Our research identifies some security risk factors associated with IoT for older adults, such as cyberattacks, lack of awareness or ignorance, inadequate security measures like weak passwords~\cite{Hirak_password_2021} and unencrypted storage, weak authentication measures~\cite{zavalyshyn_smart_2021} and physical security. Frik et al.~\cite{Frik_Privacy_2019} found that older adults expressed concerns about widespread personal data collection via IoT. These concerns encompassed issues such as intrusive monitoring, tracking, potential misuse of private information~\cite{frik_info_2023}, and a lack of transparency in data collection by these devices. The risks associated with IoT use among older adults are heightened due to several common factors, including low technical ability, physical limitations affecting security practices, dependence on potentially insecure public networks, and increased vulnerability to social engineering tactics. 

Furthermore, establishing standardized legislation and policies specifically for IoT used by older adults is essential to enhance their security and privacy.
Philip et al. argue that many industrial and consumer IoT products are designed without sufficient consideration for security and privacy~\cite{philip_internet_2021}. Thus, while adopting the rapidly expanding range of IoT, particularly those that grant older adults greater independence, it is crucial to assess the associated risks carefully. Elahi's paper emphasizes the importance of IoT Android app privacy in Ambient Assistive Living, as these apps control smart devices and require ongoing development of algorithms to improve their privacy~\cite{elahi_human-centered_2021}. Varghese et al. propose incorporating a ``Security Rating" into online product listings to enable buyers to make better-informed decisions based on customer reviews~\cite{varghese_framework_2018}.

\begin{figure*}[htbp!]
\centering
\includegraphics[width=0.7\linewidth]{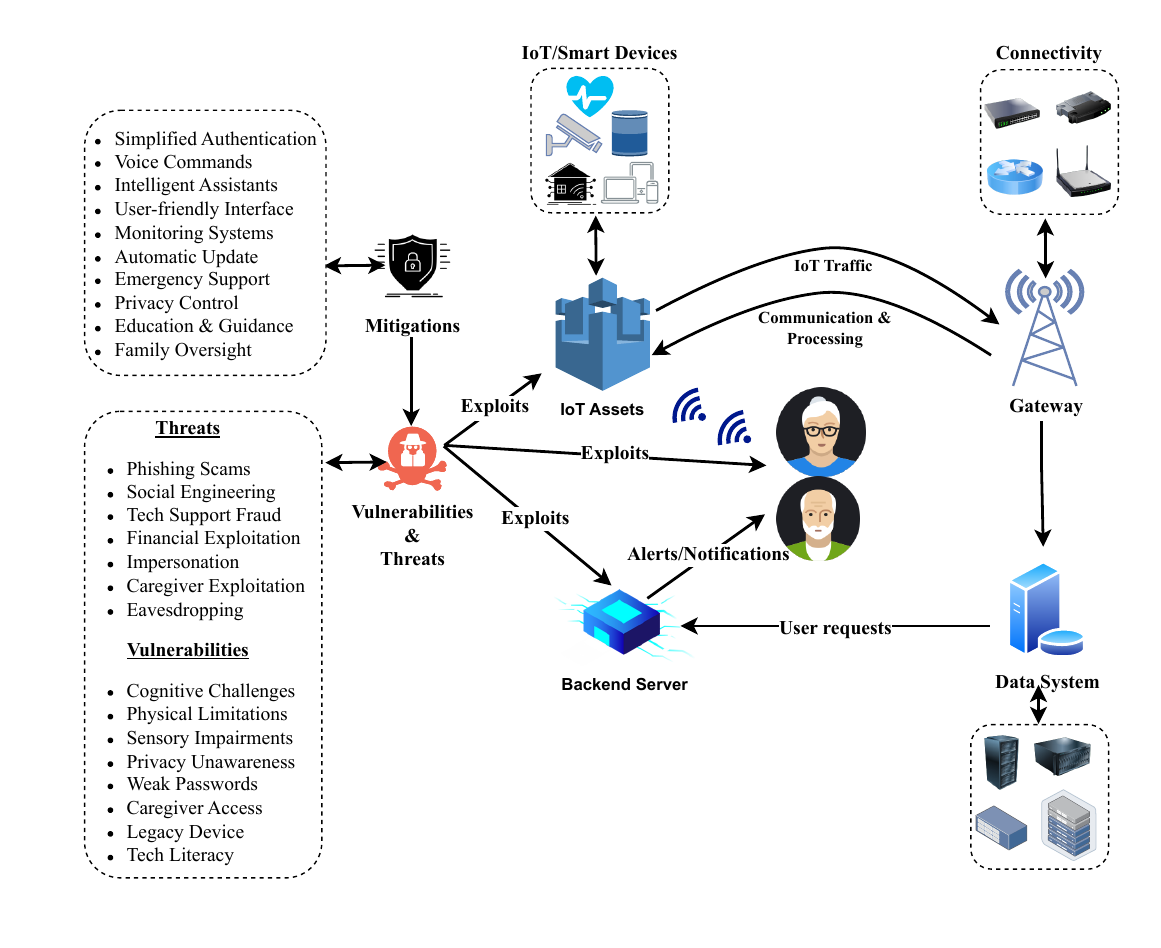}
\caption{IoT Threat Model for Older Adults derived from 44 peer-reviewed studies (2004–2024).}
\Description{A structured threat model showing IoT assets, network communication, backend systems, vulnerabilities, exploit vectors, and mitigations tailored for older adults.}
\label{fig:threat}
\end{figure*}

\subsection{IoT Risk Assessment (RQ2)}

To evaluate the current state of IoT security, privacy, accessibility, and usability for older adults, we used our proposed SPAU-IoT Framework to analyze the collected papers. Each research paper was carefully reviewed to determine if it fully, partially, or did not address the criteria within the four dimensions of the framework and the results are shown in~\autoref{tab:study_type}. 

The legend uses symbols where a solid circle (\yes) represents a full discussion of a criterion, a hollow circle ($\circ$) signifies partial discussion, and a dash (\no) indicates no discussion. As shown in the table, most of the studies concentrated on selected criteria, and none of them were able to exhaust our list of criteria. For the set of Security criteria, some studies tried to address resilience against cyber threats (S1), secure authentication mechanisms (S3), and data encryption (S4), while discussions on secure software updates (S7), incident response (S8), and guardianship (S9) for IoT tailored to older adults were very scarce. For Privacy, many papers highlighted the necessity of data minimization (P1) and ensuring user consent and control (P2), but there were fewer mentions of privacy-preserving analytics (P3), policies for data retention and disposal (P4), and the importance of privacy impact assessments (P6) for IoT for older adults. For papers that explored Accessibility, more attention was given to compatibility with assistive technology (A2) and universal accessibility (A3), however, compliance with disability act (A1), compliance with accessibility standards (A4), and error forgiveness (A5) received less or no discuss. Lastly, in the Usability dimension, the emphasis was on guided interaction (U1), clear system feedback (U2), integrated assistance (U3) and consistent user experience (U4), with just a partial talk on progressive learning path (U5) for older adults using IoT. The evaluation of the research papers through the SPAU-IoT Framework shows that although the academic community are trying as much to explore the security, privacy, accessibility, and usability aspects of IoT for older adults, there are still significant gaps within the field. Our framework provides a structural approach to identify gaps and highlights the need for more detailed research that includes all four dimensions and their associated criteria. 

The application of the SPAU-IoT Framework to a series of IoT papers focused on older adults has provided insightful findings into the current implementation of Security, Privacy, Accessibility, and Usability in these technologies. Specific papers evaluated include but are not limited to works by~\cite{alkhatib_privacy_2018}, \cite{garg_privacy_2014}, and \cite{Maresova2020}, each offering perspectives on IoT solutions for older adults. The security evaluation revealed varied implementations of security features that could help strengthen the design of IoT. For instance, Garg et al. excelled in implementing data encryption to protect sensitive user data, a crucial feature for older adults handling personal health information \cite{garg_privacy_2014}. However, most systems such as those discussed in Alkhatib's teamwork, lacked security features like resilient-to-cyber-threats and secure authentication mechanisms, highlighting a significant gap in ensuring older adults' security needs are met \cite{alkhatib_privacy_2018}. Privacy evaluation also showed inconsistency. as demonstrated in robust user-consent protocols, effectively allowing older adults to control their data \cite{Maresova2020, Mikusz2019SupportingOA}. 

In contrast, the study by Elkahlout et al. pointed out deficiencies in privacy-preserving analytics, where personal data could potentially be compromised~\cite{elkahlout_iot-based_2020}. This inconsistency across platforms underscores the ongoing challenge of implementing privacy measures in IoT solutions for older adults. Accessibility findings were generally underwhelming. The work by Turjamaa et al. highlighted a lack of compatibility with assistive technologies, which are vital for many older adults. On a positive note, Turjamaa et al. incorporated features compliant with the Americans with Disabilities Act, making their IoT solution more accessible to older adults with disabilities~\cite{Turjamaa2019}. Despite these efforts, there remains a widespread need for universally accessible designs in IoT for older adults. Usability varied significantly across the evaluated research work. Garg presented a system with guided interactions that help older adults through simplified command processes, enhancing the user experience~\cite{garg_privacy_2014}. Conversely, the system reviewed in \cite{alkhatib_privacy_2018} lacked a consistent user experience, demonstrating the need for standardization in user interface design across IoT platforms. The integrated assessment using the SPAU-IoT Framework has been instrumental in highlighting the interconnectedness of security, privacy, accessibility, and usability. Enhancements in one area could potentially improve other aspects, thereby providing a more holistic benefit to older adults. For example, improving usability by implementing guided interactions, as seen in Garg et al. work can also enhance accessibility for users with cognitive impairments~\cite{garg_privacy_2014}.

\subsubsection{\textbf{IoT Threat Model:}}
\label{sec:threat_model}

From our systematic review of $44$ peer-reviewed studies, we developed a multi-layer threat model capturing the end-to-end security and privacy risk surface in IoT deployments involving older adults. The architecture comprises four interdependent components: IoT assets such as smart home hubs, wearable health monitors, emergency alert devices, and assistive robotics; the connectivity layer including Bluetooth LE, Zigbee, Z-Wave, Wi-Fi, and cellular links often routed through home gateways; backend data systems encompassing cloud-based storage, FHIR-compliant healthcare APIs, analytics pipelines, and device management dashboards; and end users, where older adults are the primary stakeholders, sometimes operating devices independently and other times in collaboration with caregivers. Many of these assets run on constrained microcontrollers with 8–32 bit architectures, sub-512~KB RAM, and lightweight OS platforms such as FreeRTOS or Zephyr, which limits the feasibility of computationally expensive cryptographic operations.

The literature reveals $27$ recurring vulnerabilities. Weak authentication credentials appear in $68\%$ of the studies, insecure wireless channels lacking link-layer encryption in $54\%$, plaintext or weakly encrypted health data storage in $39\%$, reliance on manufacturer default configurations in $36\%$, and insufficient access control in caregiver-assisted usage scenarios in $31\%$. Other significant issues include firmware update channels without integrity verification ($28\%$), susceptibility to phishing and social engineering ($25\%$), and non-compliance with WCAG~2.1 accessibility guidelines ($18\%$). These vulnerabilities often co-occur, amplifying exploit feasibility. Attack paths emerge at three distinct layers: at the asset level, adversaries exploit side channels on wearable sensors, abuse unprotected debug interfaces such as UART or JTAG, or physically tamper with devices; at the network level, ARP spoofing, replay attacks on Zigbee clusters, Bluetooth LE key cracking, and Wi-Fi KRACK-style reinstallation attacks appear frequently; at the backend level, attackers exploit misconfigured cloud access control lists, perform SQL injection on poorly validated API endpoints, or escalate privileges in caregiver accounts.

The mitigations synthesized from the literature and security engineering best practices include enforcing TLS~1.3 with AEAD ciphers such as AES-GCM or ChaCha20-Poly1305 for all communications, applying AES-256-GCM for local storage encryption, and using HMAC-SHA256 for integrity verification. Authentication should be strengthened through multi-factor mechanisms adapted for accessibility, such as tactile hardware tokens or voice biometrics, with credentials hashed using PBKDF2 or Argon2. Secure boot mechanisms combined with cryptographically signed firmware (RSA-2048 or Ed25519) ensure update integrity, while role-based access control with time-limited delegation tokens mitigates caregiver exploitation risks. Privacy dashboards, opt-in/opt-out toggles, and fine-grained third-party access settings enable transparent user control over data flows, and on-device anomaly detection via TinyML models provides early warning for behavioral anomalies. The model is fully compatible with STRIDE and LINDDUN methodologies, enabling structured threat enumeration and privacy impact assessment. Each vulnerability can be assigned a likelihood and impact score, yielding a quantitative prioritization framework. For example, weak credentials in systems with remote access capabilities have an average CVSS~v3.1 base score of 8.2 across analyzed cases, indicating critical severity. By explicitly linking older-adult-specific human-factor risks with precise technical attack surfaces and countermeasures, this threat model delivers a security-by-design blueprint that can be directly integrated into IoT development lifecycles while supporting measurable, repeatable risk assessment.

\subsection{User-Centered Design Strategies (RQ3)}

\subsubsection{\textbf{Empirical Studies Involving Older Adults:}}
Our analysis revealed that only three papers~\cite{elahi_human-centered_2021, Frik_Privacy_2019, pal2019} from our corpus conducted User Studies. Elahi et al.~\cite{elahi_human-centered_2021} argued for a shared responsibility approach among older adults regarding privacy concerns in IoT Android apps. They propose an algorithm aimed at enhancing privacy within Ambient Assistive Living (AAL). Their methodology involved a case study where $20$ participants were invited to assess a popular chat app, with the app's description concealed. Participants individually evaluated and granted permissions on this anonymous app. Through statistical analysis, including the consideration of subjective weights, the authors demonstrated the algorithm's non-invasive nature and suitability for AAL Android users. Recognizing that apps associated with IoT play a pivotal role as control and command systems, ensuring their security is of paramount importance. In contrast, Frik et al.~\cite{Frik_Privacy_2019} conducted an interview study involving $46$ older adults to uncover common threats, concerns and misconceptions surrounding new technologies such as wearables, voice assistants, and monitoring systems. This study unveiled a spectrum of security and privacy concerns among older adults, ranging from worries about surveillance to concerns about data sharing and loss of control over personal information~\cite{frik_info_2023}. Consequently, further empirical research is needed to validate and apply these findings to IoT adoption and usage by older adults.

\subsubsection{\textbf{Critical Appraisal of the Empirical Study:}}
The study conducted by Elahi et al.~\cite{elahi_human-centered_2021} was valuable for understanding how older adults utilize IoT and the potential privacy implications of a popular chat app. The research provides real-world validity by directly exploring the perspectives of older adults and assessing how intrusive chat apps were to their privacy. However, there are limitations, such as a small sample size, focus on just one specific app and a potential lack of diversity in the participants, which may impact generalizability.  The conclusions derived mostly from technical analysis and prototypes need to be supported by more extensive empirical research on IoT security and privacy including larger samples of older adults and a variety of devices. Technical analysis and prototype evaluations often focus on one or two dimensions such as security or privacy concerns in IoT, neglecting the combination of all four: security, privacy, accessibility, and usability. These aspects are vital for practical everyday use and have an impact on adoption and effectiveness, especially among older adults. However, empirical research has also overlooked these critical factors.
For instance, Frik et al.~\cite{Frik_Privacy_2019} provide good illustrations of security and privacy concerns raised by older adults regarding new technology like wearables, smart speakers, and house assistants. Their study investigated the viewpoints of older adults regarding security and privacy, but it also overlooked how accessible and user-friendly these devices are for older adults.

\subsubsection{\textbf{Challenges for Older Adults in IoT:}}
Our study confirms that age-related impairments, such as cognitive decline, sensory impairments, and physical limitations, create significant security, privacy, accessibility, and usability challenges for older adults using IoT~\cite{poyner_privacy_2018, elahi_human-centered_2021}. The absence of all-in-one industry standards and best practices for IoT security, privacy, accessibility, and usability policies~\cite{poyner_privacy_2018} further worsens these difficulties, making it challenging to devise tailored solutions. Rapid technological advancements and deployments, along with older adults' limited technological knowledge, intensify the situation. Older adults express serious privacy concerns regarding IoT usage~\cite{garg_privacy_2014, alkhatib_privacy_2018}, as these devices often collect and store sensitive personal data, including health information. Securely storing this data and enabling traceability of access activities for auditing purposes is important as stated by Fang et al.~\cite{fang_nonintrusive_2021}. Furthermore, the acceptance of IoT monitoring systems by older adults depends on factors like data integrity~\cite{guizani_iot_2020} and personal information authenticity. Assuring them that their personal information is secure and protected is necessary before they trade off their privacy for monitoring benefits~\cite{garg_privacy_2014}.

Additionally, concerns revolve around older adults' understanding and operation of IoT, emphasizing the importance of usability and accessibility. Often, IoT design neglects the specific needs and preferences of older adults, exposing them to security risks and usability challenges. This underscores the necessity for a user-centered approach to IoT design, tailored to the requirements of this demographic. Our threat model and analysis identified vulnerabilities specific to older adults, such as cognitive challenges, physical limitations, sensory impairments, privacy unawareness, weak password practices, and the potential for caregiver access misuse. These vulnerabilities could lead to unauthorized access, accidental misconfigurations, or privacy breaches if not properly mitigated through tailored security measures and accessibility considerations. To address some of these, we propose using reliability metrics to assess a multi-modal biometric authentication involving face, voice, and pattern recognition for IoT access. Another major barrier to secure IoT use is the lack of defined legislation and standards surrounding IoT design for this demographic. The industry must prioritize creating and enforcing best practices, standards, and regulations for IoT design to protect older adults.

\section{Implications}

\subsection{Security-by-Design for Older Adults}
Our analysis revealed that $68\%$ of reviewed systems relied on weak or static authentication credentials, often due to the need for memorable passwords, and $54\%$ employed insecure wireless protocols such as unencrypted Bluetooth or Zigbee channels. For example, in Ganesh et al.'s wearable fall detection system, default PIN-based authentication was retained throughout deployment, creating a predictable attack surface~\cite{ganesh_iot-based_2019}. Similarly, Pinto et al.'s We-Care wristband initially transmitted health telemetry over unencrypted Zigbee links~\cite{pinto_we-care_2017}, making passive interception feasible with low-cost hardware. These cases underscore the necessity of embedding strong security controls during the IoT design lifecycle rather than applying reactive patches after deployment. Recommended measures include default-enabled multi-factor authentication, elliptic-curve-based device pairing to secure onboarding, and TLS~1.3 for all communications. To address memory and dexterity constraints without weakening security, controls should use adaptive authentication such as fingerprint or voice verification to reduce cognitive load while resisting brute-force and spoofing attacks~\cite{chauhan2018secure,almohamade2022continuous}.

\subsection{Accessibility as a Security Enabler}
$31\%$ of systems lacked fine-grained caregiver access controls, and over a quarter failed to comply with WCAG~2.1 accessibility standards. In some deployments, such as Fang et al.'s smart home motion-tracking system~\cite{fang_nonintrusive_2021}, the absence of configurable caregiver permissions led to shared credentials between users and caregivers, creating unnecessary privilege escalation risks. Accessibility in this context is not only an inclusivity requirement but also a security enabler: adaptable visual, auditory, and cognitive interfaces can significantly reduce the risk of misconfiguration and accidental exposure of sensitive data. For example, a WCAG-compliant interface with large-font displays, high contrast themes, and auditory feedback can prevent menu selection errors that inadvertently disable encryption or open network ports. Integrating the SPAU–IoT framework into secure coding practices ensures that usability features, such as simplified menus, step-by-step configuration wizards, and context-aware error prompts are implemented without introducing new attack vectors or bypassing security controls.

\subsection{Edge-Intelligent Gateways}
Our multi-layer threat model indicates that older adults' IoT environments often rely on commodity gateways that provide minimal security validation, leaving devices exposed to replay attacks, ARP spoofing, and backend compromise. In Pinto et al.'s We-Care system~\cite{pinto_we-care_2017}, health telemetry was sent directly to the cloud without intermediate validation, enabling potential injection attacks from compromised devices. We recommend the deployment of edge-intelligent gateways capable of local anomaly detection, automated and code-signed firmware updates, and fine-grained caregiver role enforcement. These gateways can serve as both a security enforcement point and an accessibility facilitator, verifying configurations before deployment, detecting deviations in behavioral patterns (e.g., abnormal device usage times), and providing assisted troubleshooting tailored to older-adult cognitive and sensory profiles~\cite{nandikotkur2023seniorsentry,tougay2019secure,hu2022gatekeeper}.

\subsection{Regulatory Alignment and Standardization}
We found that few studies explicitly aligned their designs with GDPR, HIPAA, or NIST IoT cybersecurity guidelines, despite many handling highly sensitive health and behavioral data. For example, Philip et al.'s vital signs monitoring platform~\cite{philip_internet_2021} collected continuous blood pressure and heart rate data without documenting data retention or encryption policies, leaving compliance gaps with HIPAA's security rule. Mapping the SPAU–IoT criteria to existing legal frameworks enables developers to meet both technical performance and policy standards while ensuring privacy-by-design. This mapping enables interoperable IoT solutions to scale across aging populations while preserving cryptographic strength, access control, and audit-ready privacy.

\section{Limitations and Future Work}
Our search strategy may have missed some HCI papers on older adults' interactions with connected technologies when the title or abstract does not explicitly mention ``IoT'' or SPAU, since we required these terms for reproducibility across databases. This review also draws on $44$ studies published between $2004$ and $2024$ and excludes non-English publications. Combined with the scarcity of reproducible experiments and detailed accessibility evaluations, limit cross-study comparison and quantitative meta-analysis. To address these gaps, we are collaborating with gerontology and IoT security institutions to conduct empirical studies involving real-world deployments, adversarial testing, and longitudinal monitoring, and we will use these to validate and refine the SPAU IoT framework and threat model in operational settings.

\section{Conclusion}  
We analyzed $44$ peer-reviewed studies ($2004$-$2024$) on IoT for older adults, examining the intersection of security, privacy, accessibility, and usability in this demographic's technology adoption. We found that $86\%$ of deployments target remote health monitoring and smart home automation, incorporating physiological sensing, fall detection, medication reminders, and caregiver-assisted operation. Quantitative analysis identified older-adult-specific vulnerabilities: weak or static authentication credentials in $68\%$ of systems, insecure Bluetooth/Zigbee channels in $54\%$, unencrypted health data storage in $39\%$, reliance on default configurations in $36\%$, and inadequate caregiver access control in $31\%$. The SPAU-IoT framework uses biometric and voice authentication for users with memory impairments, WCAG-compliant adaptive interfaces for sensory limitations, and fine-grained caregiver permissions that preserve autonomy. The threat model maps asset, network, and backend vulnerabilities to attacks such as phishing of health alerts, Zigbee replay, ARP spoofing on home Wi-Fi, and caregiver impersonation. Mitigations deploy TLS~1.3 communication, AES-256-encrypted storage, elliptic-curve device pairing, signed firmware updates, and on-device anomaly detection, yielding a reproducible security-by-design blueprint that protects older adults' data and independence against advanced adversaries.

\section*{Acknowledgment}
We would like to acknowledge the Das Agency and Security (DAS) Lab at George Mason University and the Inclusive Security and Privacy-focused Innovative Research in Information Technology (INSPIRIT) Lab at the University of Denver. The opinions expressed in this work are solely those of the authors.

\bibliographystyle{ACM-Reference-Format}
\bibliography{ASIACCS2026}

\appendix
\section*{APPENDIX }
\label{codebook}

\section{Codebook and Coding Procedure}

This appendix provides the complete codebook used during full-text
screening and thematic analysis. Two researchers independently coded an initial subset of papers using a hybrid approach that combined codes derived from the SPAU with codes that emerged during reading of the corpus. Initial inter-rater agreement was 91\% for title and abstract screening and 89\% for full-text screening. All disagreements were resolved through discussion and consensus. The final codebook reflects the SPAU criteria, the thematic domains, and the methodological attributes extracted during data collection. Codebook Used for Thematic Analysis is shown in~\autoref{tab:codebook}.
The codes were applied iteratively across all 44 papers in the final corpus. Codes such as \emph{Security Measures}, \emph{Privacy Concerns}, \emph{Accessibility Features}, and \emph{Usability Challenges} correspond to the SPAU dimensions. Thematic codes such as \emph{Healthcare Applications}, \emph{Independent Living},
\emph{Legislation and Policy}, and \emph{Caregiver Involvement} emerged from patterns observed during full-text screening. Methodological codes including \emph{Study Type}, \emph{IoT Device Type}, \emph{Target Population}, and \emph{Evaluation Method} supported the descriptive analysis.
Table~\ref{tab:included-papers} lists all 44 papers included in our final corpus, along with basic descriptive metadata. Study type and IoT device type are derived from the full-text screening sheet.

\begin{table*}
\centering
\caption{Codebook Used for Thematic Analysis}
\begin{tabular}{p{3.5cm} p{5cm} p{5.2cm}}
\toprule
\textbf{Code} & \textbf{Definition} & \textbf{Example Excerpt} \\
\midrule

Healthcare Applications &
Studies describing IoT systems for health monitoring, chronic disease management, emergency detection, rehabilitation, or vital sign tracking. & Wearable sensors track vital signs and provide notification. \\

Independent Living & Research focused on supporting daily routines, fall detection, mobility, environmental monitoring, or home automation for older adults. & Motion sensors monitor activity to detect deviations in routine. \\

Legislation and Policy & Papers addressing regulatory, ethical, governance, or policy issues related to IoT data management and aging populations. & Data retention regulations create uncertainty for long-term sensor deployments. \\

Security Measures & Mentions of authentication, encryption, access control, secure update mechanisms, or threat mitigation strategies. These codes correspond to the nine security criteria (S1–S9) in the 
SPAU–IoT Framework. & The device uses a single password without multi-factor authentication. \\

Privacy Concerns & Issues related to data collection, sharing, consent, surveillance, or user awareness of data practices. These codes correspond to the eight privacy criteria (P1–P8) in the 
SPAU–IoT Framework. & Participants were unsure who could access their activity data. \\

Accessibility Features & Accommodations for sensory, motor, or cognitive limitations in IoT interfaces or device interactions. These codes correspond to the five accessibility criteria (A1–A5) in the 
SPAU–IoT Framework. &Large buttons and simplified menus assist users with limited dexterity. \\

Usability Challenges & Difficulties related to configuration, learnability, feedback clarity,
error recovery, or task complexity. These codes correspond to the five usability criteria (U1–U5) in the SPAU–IoT Framework. & Users struggled to interpret device alerts due to unclear interface feedback. \\

Caregiver Involvement & Cases where caregivers assist with configuration, monitoring, access delegation, or device decision-making. &
 Caregivers used shared credentials to adjust device settings. \\

Study Type & Classification of the study as technical analysis, prototype evaluation, 
or user study & The paper reports a prototype evaluation without end-user testing. \\

IoT Device Type & Type of device examined: wearables, environmental sensors, smart home systems, medical IoT, or mixed deployments. &
 The system integrates a smart wristband with home motion sensors. \\

Target Population & Demographic characteristics of participants, especially older adults (65+) & The evaluation involved adults aged 65 and above. \\

\bottomrule
\end{tabular}
\label{tab:codebook}
\end{table*}


\begin{table*}[ht]
\centering
\caption{Included Papers and Metadata (n = 44)}
\begin{tabular}{p{0.2cm} p{6.0cm} p{0.9cm} p{3.2cm} p{2.8cm} p{2.6cm}}
\toprule
\textbf{ID} & \textbf{Title} & \textbf{Year} & \textbf{Venue / Source} &
\textbf{Study Type} & \textbf{IoT Device Type} \\
\midrule

1 & Privacy and the Internet of Things (IoT) Monitoring Solutions for Older Adults: A Review & 2018 & HIC (IOS Press) & Technical analysis & Monitoring \\

2 & A Performance Analysis of an IoT-aware Elderly Monitoring System & 2018 & SpliTech & Prototype evaluation & Monitoring \\

3 & IoT-Based Healthcare and Monitoring Systems for the Elderly: A Literature Survey Study & 2020 & iCareTech & Technical analysis & Monitoring \\

4 & Privacy concerns in assisted living technologies & 2013 & Annals of Telecom & Technical analysis & Monitoring \\

5 & Towards Privacy-Preserved Aging in Place: A Systematic Review & 2021 & Sensors (MDPI) & Technical analysis & Monitoring; Care \& Support \\

6 & IoT Healthcare Monitoring Systems Overview for Elderly Population & 2020 & IWCMC & Technical analysis & Monitoring \\

7 & Trust, Privacy and Security, and Accessibility Considerations When Conducting Mobile Technologies Research With Older Adults & 2020 & NASEM Workshop Proceedings & Technical analysis & Monitoring \\

8 & Illuminating Privacy and Security Concerns in Older Adults’ Technology Adoption & 2024 & Work, Aging and Retirement (OUP) & Technical analysis & Monitoring; Care \& Support \\

9 & A review of IoT systems to enable independence for the elderly and disabled individuals & 2023 & Internet of Things (Elsevier) & Technical analysis & Care \& Support \\

10 & Internet of Things for In-Home Health Monitoring Systems: Current Advances, Challenges and Future Directions & 2021 & IEEE JSAC & Technical analysis & Monitoring \\

11 & Privacy and security of consumer IoT devices for the pervasive monitoring of vulnerable people & 2018 & Cybersecurity of the IoT & Technical analysis & Monitoring \\

12 & Privacy Versus Autonomy: A Tradeoff Model for Smart Home Monitoring Technologies & 2011 & IEEE EMBS & Technical analysis & Monitoring; Care \& Support \\

13 & Smart Home Care: Towards Supporting Elderlies in the Comfort and Safety of their (Smart) Homes & 2021 & LADC & Technical analysis & Care \& Support \\

14 & Vision: Stewardship of Smart Devices Security for the Aging Population & 2021 & EuroUSEC (ACM) & Technical analysis & Monitoring \\

15 & Supporting Older Adults Using Privacy-Aware IoT Analytics & 2018 & Living in the Internet of Things & Prototype evaluation & Monitoring; Care \& Support \\

16 & Design and Development of a User-Friendly, Smart, and Assistive Kitchen for the Mauritian Elderly & 2023 & SmartTechCon (IEEE) & Prototype evaluation & Monitoring;  Care \& Support \\

17 & The smart home for the elderly:
Perceptions, technologies and psychological accessibilities: The requirements analysis for the elderly in Thailand & 2018 & IEEE ICAT & Prototype  & Monitoring; Detection; Care \& Support \\

18 & Designing an Intelligent UI/UX System Based on the Cognitive Response for Smart Senior & 2016 & ICSITech (IEEE) & Technical analysis & Monitoring; Care \& Support \\

19 & How can smart home help “New elders” aging in place and building connectivity & 2020 & Intelligent Environments (IE) & Technical analysis & Monitoring; Detection \\

20 & Usability Evaluation Framework of Smart Home Applications for Senior Citizens & 2022 & ICSTE (IEEE) & Technical analysis & Monitoring \\

21 & Health–Related ICT Solutions of Smart Environments for Elderly - Systematic Review & 2020 & IEEE Access & Technical analysis & Monitoring; Care \& Support \\

22 & FEEL: Federated Learning Framework for Elderly Healthcare Using Edge-IoMT & 2023 & IEEE Trans. Computational Social Systems & Prototype evaluation & Monitoring; Detection; Care \& Support \\

\bottomrule
\end{tabular}
\label{tab:included-papers}
\end{table*}

\begin{table*}[ht]
\centering
\begin{tabular}{p{0.2cm} p{6.0cm} p{0.9cm} p{3.2cm} p{2.8cm} p{2.6cm}}
\toprule
\textbf{ID} & \textbf{Title} & \textbf{Year} & \textbf{Venue / Source} &
\textbf{Study Type} & \textbf{IoT Device Type} \\
\midrule

23 & IoT-Enabled Community Care for Ageing-in-Place: The Singapore Experience & 2020 & EECSI  & Technical analysis & Monitoring; Care \& Support \\

24 & AAL Platforms Challenges in IoT Era: A Tertiary Study & 2018 & IEEE SoSE & Technical analysis & Monitoring; Care \& Support \\

25 & Smart Homes and Quality of Life for the Elderly: A Systematic Review & 2017 & IEEE ISM  & Technical analysis & Monitoring \\

26 & Fundamental Issues in the Design of Smart Home for Elderly Healthcare & 2019 & ICSAI (IEEE) & Technical analysis & Monitoring; Care \& Support \\

27 & How smart homes are used to support older people: An integrative review & 2019 & Int. J. Older People Nursing & Technical analysis & Monitoring;  Care \& Support \\

28 & Ten questions concerning smart and healthy built environments for older adults & 2023 & Building and Environment & Technical analysis & Monitoring; Care \& Support \\

29 & SMART BEAR: A Large-Scale Pilot Supporting the Independent Living of the Seniors in a Smart Environment & 2020 & IEEE EMBC & Technical analysis & Monitoring; Detection; Care \& Support; Navigation \\

30 & Designing Digital Technologies and Safeguards for Improving Activities and Well-Being for Aging in Place & 2020 & HCII (Springer LNCS) & Technical analysis & Monitoring; Navigation; Care \& Support \\

31 & Improving the Quality of Life for Older People: From Smart Sensors to Distributed Platforms & 2019 & IEEE CSCS & Technical analysis & Monitoring; Care \& Support \\

32 & A Nonintrusive Elderly Home Monitoring System & 2021 & IEEE Internet of Things Journal & Prototype evaluation & Monitoring \\

33 & Smart Bed for Health Monitoring System Using IoT & 2019 & IC3I  & Prototype evaluation & Monitoring; Detection \\

34 & An IoT-Aware Approach for Elderly-Friendly Cities & 2018 & IEEE Access & Technical analysis & Monitoring;  Care \& Support \\

35 & IoT \& Cloud Computing-based Remote Healthcare Monitoring System for an Elderly-Centered Care & 2021 & EHB (IEEE) & Prototype evaluation & Monitoring; Care \& Support \\

36 & We-Care: An IoT-based Health Care System for Elderly People & 2018 & IEEE ICIT & Prototype evaluation & Monitoring; Detection; Care \& Support \\

37 & An Updated Watch-over System Using an IoT Device, for Elderly People Living by Themselves & 2019 & ICSR & Prototype evaluation & Monitoring; Care \& Support \\

38 & Usability in the App Interface Designing for the Elderly with Low-Vision in Taiwan and Thailand & 2020 & IEEE ECBioMed  & Technical analysis & Monitoring; Care \& Support \\

39 & IoT-based Intelligent Building System for Community-based Care & 2022 & ICCE–TW  (IEEE) & Technical analysis & Monitoring; Care \& Support \\

40 & Designing Smart Home Controls for Elderly & 2020 & PETRA (ACM) & Technical analysis & Monitoring; Care \& Support \\

41 & IoT-based systems for improving older adults’ wellbeing: a systematic review & 2019 & CISTI & Technical analysis & Monitoring; Detection; Care \& Support \\

42 & A human-centered artificial intelligence approach for privacy protection of elderly App users in smart cities & 2021 & Neurocomputing & User study & Monitoring; Detection; Care \& Support \\

43 & Privacy and Security Threat Models and Mitigation Strategies of Older Adults & 2019 & SOUPS (USENIX) & User study & Monitoring; Care \& Support \\

44 & Embracing the Smart-Home Revolution in Asia by the Elderly: An End-User Negative Perception Modeling & 2019 & IEEE Access & User study & Monitoring; \\

\bottomrule
\end{tabular}
\end{table*}

\end{document}